\documentclass[12pt]{article}
\usepackage[utf8]{inputenc}
\usepackage[DIV=13]{typearea}
\usepackage{ragged2e}
\usepackage{calligra,amsmath,amsfonts,bbm,mathrsfs,amssymb}
\usepackage{slashed,cancel}
\usepackage{cite}
\usepackage{bm} 
\usepackage{hyperref}
\hypersetup{colorlinks=true,urlcolor=magenta,anchorcolor=blue,citecolor=blue,filecolor=blue,
            linkcolor=magenta,menucolor=blue, linktocpage=true,pdfproducer=medialab}
\usepackage[english]{babel}
\usepackage{indentfirst}
\usepackage{graphicx}
\usepackage{float}
\usepackage{enumerate}
\usepackage{caption}
\usepackage{subfigure}
\usepackage{multirow}
\usepackage[normalem]{ulem} 
\usepackage{booktabs}
\usepackage{physics}

\newcommand{\email}[1]{\href{mailto:#1}{\tt #1}}

\numberwithin{equation}{section}

\newcommand{\blue}[1]{\color{blue} #1 \color{black}}

\newcommand{\magenta}[1]{\color{magenta} #1 \color{black}}
\newcommand{\be}{\begin{equation}}
\newcommand{\ee}{\end{equation}}
\newcommand{\ba}{\begin{equation}\begin{aligned}}
\newcommand{\ea}{\end{aligned}\end{equation}}

\newcommand{\rlb}{\left(}
\newcommand{\rrb}{\right)}
\newcommand{\slb}{\left[}
\newcommand{\srb}{\right]}

\newcommand{\sL}{\mathscr{L}}

\newcommand{\hc}{\text{h.c.}}
\newcommand{\nn}{\nonumber}
\renewcommand{\vev}[1]{\langle #1\rangle}

\newcommand{\ov}[1]{\overline{#1}}

\newcommand{\TeV}{\ \text{TeV}}
\newcommand{\GeV}{\ \text{GeV}}

\def\Re{{\texttt{Re}}}
\def\Im{{\texttt{Im}}}
\def\vs{{\textit vs.} }

\newcommand{\gl}{\lambda}
\newcommand{\tgl}{\widetilde{\lambda}}
\newcommand{\tm}{\widetilde{m}}

\newcommand\subsetsim{\mathrel{%
  \ooalign{\raise0.2ex\hbox{$\subset$}\cr\hidewidth\raise-0.8ex\hbox{\scalebox{0.9}{$\sim$}}\hidewidth\cr}}}
\newcommand*{\diff}[1]{\text{d}#1}

\begin{document} 
\renewcommand*{\thefootnote}{\fnsymbol{footnote}}
\begin{titlepage}

\vspace*{-1cm}
\flushleft{\magenta{IFT-UAM/CSIC-23-36}}
\\[1cm]

\begin{center}
\blue{\bf\LARGE\boldmath Flavour and Higgs physics  in }\\[2mm] 
\blue{\bf\LARGE $Z_2$-symmetric 2HD models}\\[2mm] 
\blue{\bf\LARGE near the decoupling limit}\\[4mm]
\unboldmath
\centering
\vskip .3cm
\end{center}
\vskip 0.5  cm
\begin{center}
{\large\bf Arturo de Giorgi}${}^{a)}$~\footnote{\email{arturo.degiorgi@uam.es}},
{\large\bf Fotis Koutroulis}${}^{b)}$~\footnote{\email{fotis.koutroulis@fuw.edu.pl}},\\
\vskip .4cm
{\large\bf Luca Merlo}${}^{a)}$~\footnote{\email{luca.merlo@uam.es}}, and 
{\large\bf Stefan Pokorski}${}^{b)}$~\footnote{\email{Stefan.Pokorski@fuw.edu.pl}}
\vskip .7cm
{\footnotesize
${}^{a)}$ Departamento de F\'isica Te\'orica and Instituto de F\'isica Te\'orica UAM/CSIC,\\
Universidad Aut\'onoma de Madrid, Cantoblanco, 28049, Madrid, Spain
\vskip .2cm
${}^{b)}$ Institute of Theoretical Physics, Faculty of Physics,\\ 
University of Warsaw, Pasteura 5, PL 02-093, Warsaw, Poland
}
\end{center}
\vskip 2cm
\begin{abstract}
\justifying
With no evidence of any exotic particle detected so far beyond the Standard Model, the new physics may lie above the presently accessible energies at colliders and, at low-energies, can be accounted for via an effective description. The interplay of flavour and Higgs physics data  allows setting stringent bounds on the parameters of the effective Lagrangian. In this paper, we focus on $Z_2$-symmetric two Higgs doublet models near the decoupling limit: the corresponding effective description relies on only a few parameters, thus predicting many interesting correlations between observables that work as tests of the theory. We present the results of a global fit to the existing data, updating and extending over the past literature.  We comment on the triple Higgs coupling as a probe of an extended scalar sector and on the recent CDF II measurement of the $W$-mass.

\end{abstract}
\end{titlepage}
\setcounter{footnote}{0}

\pdfbookmark[1]{Table of Contents}{tableofcontents}
\tableofcontents

\renewcommand*{\thefootnote}{\arabic{footnote}}
%
%
\section{Introduction}
\label{sec:intro}

After the July 4th 2012  announcement~\cite{ATLAS:2012yve,CMS:2012qbp} of the discovery of a resonance  compatible with the Standard Model (SM) Higgs boson, no other particle has been detected. On the other hand, experimental data from the last decade is being used to increase the precision of our knowledge on many different observables, resulting in very accurate indirect searches.

In the absence of a specific New Physics (NP), one approach to studying possible deviations from the SM predictions is the Standard Model Effective Field Theory (SMEFT)~\cite{Buchmuller:1985jz,Grzadkowski:2010es} that is a basis of independent operators, ordered in terms of mass dimensions, invariant under the SM gauge symmetry and constructed out of the SM fields~\footnote{As the nature of the Higgs field is still uncertain up to a certain degree, alternative effective descriptions have been proposed~\cite{Feruglio:1992wf,Grinstein:2007iv,Contino:2010mh,Alonso:2012px,Alonso:2012pz,Brivio:2013pma,Gavela:2014vra,Buchalla:2013rka,Brivio:2014pfa,Kozow:2019txg}, providing different expectations in certain observables.}.
The hope is to extract information on the ultraviolet (UV) completion by identifying specific patterns and synergies among the SMEFT operators. 

Even restricting to the subgroup with mass dimension $d=6$, the number of such operators is tremendous, 2499, considering all the flavour contractions~\cite{Alonso:2013hga}. Although it is possible to perform global fits on a reduced subset of operators under certain reasonable assumptions, no clear evidence of a {\it specific} NP has emerged (see for example Ref.~\cite{Eboli:2021unw} for a recent fit to the electroweak Higgs data). This occurs despite the rapport on some apparent anomalies in particle physics. To be mentioned is the new measurement from the CDF II collaboration of the $W$-gauge boson mass~\cite{CDF:2022hxs}, that exceeds the SM prediction by approximately $7\sigma$ (but not confirmed by the last ATLAS result~\cite{ATLAS-CONF-2023-004}): assuming that the CDF II result is correct, different explanations have been proposed to account for the deviation from the SM prediction, from tree-level modification of the muon decay induced by exotic leptons (see for example Refs.~\cite{Blennow:2022yfm,Arias-Aragon:2022ats,deGiorgi:2022xhr}), to loop quantum corrections to the two-point functions of the SM gauge bosons  (see for example Refs.~\cite{Sakurai:2022hwh,Paul:2022dds}). Another example is the charged $B$-anomalies, hinting to lepton flavour universality violation in the $B$-meson semi-leptonic decays $B\rightarrow D^{(\ast)}l\nu$: the experimental results, combining data from the BaBar~\cite{BaBar:2012obs,BaBar:2013mob} and Belle~\cite{Belle:2007qnm,Belle:2009xqm,Belle:2010tvu} collaborations, exceeds the SM prediction by $3.5\sigma$~\cite{Sakaki:2013bfa}.  Recently, the LHCb collaboration~\cite{PuthumanaillamTALK} announced its results on this observable, lowering the deviation of the world average to the $3\sigma$ level. Considering that other apparent discrepancies in the $B$-decays have been clarified by the LHCb collaboration~\cite{LHCb:2022qnv,LHCb:2022zom},  explanations of the charged $B$-anomalies in terms of exotic charged gauge bosons (see e.g. Ref.~\cite{Kumar:2018kmr}) or a charged Higgs boson (see e.g. Refs.~\cite{Iguro:2022uzz,Blanke:2022pjy,Kumar:2022rcf}) or leptoquarks (see e.g. Refs.~\cite{Calibbi:2017qbu,Heeck:2018ntp,Fornal:2018dqn}) appear much less plausible, even if still possible.

This bottom-up approach has been largely adopted in the recent literature and,  given the lack of clear indications towards a specific NP, we will abandon it here.  Instead, we will assume a concrete extension of the SM and analyse its effects in the low-energy effective Lagrangian approximation, with the aim of exploring the associated phenomenology in the reduced  NP parameter space. Our choice is the two Higgs doublet models (2HDMs)~\cite{Lee:1973iz,Glashow:1976nt,Deshpande:1977rw,Donoghue:1978cj,Inoue:1982pi,Flores:1982pr,Gunion:1984yn,Gunion:1986nh}. Although such extensions of the SM do not address the most pressing questions like the origin of the active neutrino masses, the existence of dark matter and the baryon asymmetry of the Universe, they are nevertheless a useful theoretical laboratory. They permit to investigate the room left by the present flavour and Higgs sector experimental data for NP in the scalar potential and in the effective Yukawa couplings. They are also interesting in the context of supersymmetric extensions of the SM.

Most of the analyses present in the literature in the framework of the EFT approach treat data on the electroweak (EW) and Higgs sectors separately from flavour data. Only a few examples appeared attempting to merge all the information, although at the cost of assuming specific flavour structures. This is the case of Refs.~\cite{Fuchs:2020uoc,Alonso-Gonzalez:2021jsa,Alonso-Gonzalez:2021tpo} and subsequent Refs.~\cite{Bahl:2022yrs,Brod:2022bww}, where the generic conclusion is that flavour data and EW/Higgs data are close to being comparable to each other in their sensitivity to NP effects.  In this paper, we will extend upon the study in Refs.~\cite{Alonso-Gonzalez:2021jsa,Alonso-Gonzalez:2021tpo} based on the Minimal Flavour Violation (MFV) ansatz~\cite{Chivukula:1987py,DAmbrosio:2002vsn,Cirigliano:2005ck,Davidson:2006bd,Gavela:2009cd,Alonso:2011jd} and on the Froggatt-Nielsen (FN) mechanism~\cite{Froggatt:1978nt}, two very well-known contexts in the literature, see Refs.~\cite{Grinstein:2010ve,Feldmann:2010yp,Albrecht:2010xh,Guadagnoli:2011id,Alonso:2011yg,Buras:2011zb,Redi:2011zi,Buras:2011wi,Alonso:2012jc,Alonso:2012fy,Alonso:2013mca,Lopez-Honorez:2013wla,Alonso:2016onw,Dinh:2017smk,Arias-Aragon:2017eww,Merlo:2018rin,Arias-Aragon:2020bzy,Arias-Aragon:2020qip} and~\cite{Dudas:1995yu,Chankowski:2005qp,Altarelli:2002sg,Altarelli:2012ia,Bergstrom:2014owa}, respectively.  The MFV and the FN represent two extreme cases: the fermion-Higgs couplings are flavour conserving or largely violating it, respectively; moreover, there is only one universal CP violating phase for each fermion sector in the MFV case, while several phases are present in the FN one.
Our framework is the effective Lagrangian derived from $Z_2$-symmetric versions of the 2HDMs, where the role of the $Z_2$ symmetry is to ensure  protection from tree-level flavour changing neutral current (FCNC) contributions~\cite{Glashow:1976nt}, but still allowing effects in flavour changing processes due to the presence of charged Higgses.  Thus, they represent a middle ground between the two aforementioned constructions based on flavour symmetries.
Furthermore, as we will discuss, in the effective Yukawa couplings there is only one universal CP violating phase for all fermions of the SM.

Our analysis is valid near the decoupling limit, which will be defined in Sect.~\ref{sec:Lag}, of all the heavy scalar states.
This hypothesis is motivated by the absence of any discovery of additional scalars and we will assume them to be heavy enough to justify an effective description. In consequence, all the heavy scalars are almost degenerate in mass, so  their low-energy effects are to a good approximation controlled by only one mass scale. This last condition will be relaxed when discussing the possible explanation of the CDF II measurement of the $W$ mass.

The bulk of our analysis depends on only three new NP parameters, that is the common mass of the heavier scalar degrees of freedom, one coupling in the scalar potential and $\tan\beta$, characteristic of the 2HDM constructions. The predictive power is therefore high and we make use of it providing the expected maximum deviation in the triple Higgs coupling from its SM value for each $Z_2$-symmetric 2HDM realisation.

The structure of the paper can be read in the table of Contents.

%
%
\section{The 2HDM Lagrangian}
\label{sec:Lag}

The general 2HDM and its $Z_2$ symmetric versions have been extensively discussed in the literature (see for example Refs.~\cite{Gunion:1989we,Carena:2002es,Branco:2011iw,Gorbahn:2015gxa,Brehmer:2015rna,Egana-Ugrinovic:2015vgy,Belusca-Maito:2016dqe,Dawson:2017vgm,Dawson:2022cmu}). The main motivation for imposing a  $Z_2$ symmetry on a 2HDM is to forbid tree-level FCNC contributions. For the sake of easy reference to the $Z_2$ symmetric versions and for fixing the notation used throughout the paper,  we collect here (and in the App.~\ref{App.A}) their key ingredients.

A discrete Abelian $Z_2$ symmetry quantum numbers are assigned to the two Higgs doublets $\Phi_1$ and $\Phi_2$, even and odd under $Z_2$ respectively, and to the fermion fields.  The assignment of the $Z_2$ quantum numbers identifies the {\it original} basis for those fields.  Focusing first on the scalar potential of the model, the $Z_2$ symmetry restricts its form (see App.~\ref{App.A}), and the vacuum of the model is fixed in terms of the  mass parameters and the couplings to be:
\be
\vev{\Phi_1^\dag\Phi_1}=\dfrac{v_1^2}{2}\,,\qquad
\vev{\Phi_2^\dag\Phi_2}=\dfrac{v_2^2}{2}\,,\qquad
\vev{\Phi_1^\dag\Phi_2}=\dfrac{v_{12}^2}{2}\,,\qquad
v_1^2+v_2^2=v^2\,
\ee
where $v_1$ and $v_2$ are real, $v_{12}$ is in general complex, and $v = 246 \GeV$.  
Moreover, the CP violating phase of the mixed Higgs condensate we denote as
\be
\xi\equiv \arg\left(\Phi_1^\dag\Phi_2\right)\,.
\ee

It is convenient to rewrite the scalar potential in the {\it Higgs} basis where only one of the two Higgses acquires a VEV.  Performing $U(2)$ rotations of the Higgs fields  in a two-dimensional complex plane by the angle $\beta$
defined as
\be
\tan\beta=\frac{v_2}{v_1}\,,
\ee
one obtains:
\begin{align}
    V(H_1,H_2)=\,& \widetilde{m}_1^2 H_1^\dagger H_1 +\widetilde{m}_2^2 H_2^\dagger H_2 -\rlb \widetilde{m}_{12}^2H_1^\dagger H_2 +\text{h.c.}\rrb+\dfrac{1}{2}\widetilde{\gl}_1 \rlb H_1^\dagger H_1\rrb^2+\nn\\
    &+\dfrac{1}{2}\widetilde{\gl}_2 \rlb H_2^\dagger H_2\rrb^2 + \widetilde{\gl}_3 \rlb H_1^\dagger H_1\rrb \rlb H_2^\dagger H_2\rrb + \widetilde{\gl}_4 \rlb H_1^\dagger H_2\rrb \rlb H_2^\dagger H_1\rrb+ \\
    &+\slb \dfrac{1}{2} \widetilde{\gl}_5 \rlb H_1^\dagger H_2\rrb^2  + \widetilde{\gl}_6 \rlb H_1^\dagger H_1\rrb \rlb H_1^\dagger H_2\rrb + \widetilde{\gl}_7 \rlb H_2^\dagger H_2\rrb \rlb H_1^\dagger H_2\rrb +\text{h.c.} \srb\,.\nn
\end{align}
The expressions for the tilde parameters in terms of the parameters defined in the original basis are given in the App.~\ref{App.A}. The parameters $\widetilde{m}_1^2$, $\widetilde{m}_2^2$, $\widetilde{\gl}_1$, $\widetilde{\gl}_2$, $\widetilde{\gl}_3$ and $\widetilde{\gl}_4$ are real, while $\widetilde{m}_{12}^2$, $\widetilde{\gl}_5$, $\widetilde{\gl}_6$ and $\widetilde{\gl}_7$ are in general complex.  However, as discussed in App.~\ref{App.A}, 
for the $Z_2$ symmetric models there are only two independent CP-violating phases: 
\be
\theta_1\equiv \dfrac{1}{2}\arg{(\tgl_6^2\tgl_5^\ast)}\,, \qquad\qquad 
\theta_2\equiv \dfrac{1}{2}\arg{(\tgl_7^2\tgl_5^\ast)}\,.
\ee
Enforcing the condition that 
\be
\vev{H^\dag_1H_1}=\dfrac{v^2}{2}\,,\qquad\qquad 
\vev{H^\dag_2H_2}=0\,,
\ee
leads to two relations between four of the parameters,
\be
\widetilde{m}_1^2=-\dfrac{1}{2}\tgl_1 v^2\,, \qquad\qquad \widetilde{m}_{12}^2=-\dfrac{1}{2}\tgl_6 v^2\,,
\ee
reducing the number of independent parameters.

In the vacuum defined by the scalar potential, five degrees of freedom acquire masses, while the remaining three remain massless, corresponding to the three longitudinal components of the EW gauge bosons. The two charged Higgs bosons $H^\pm$ receive a mass that reads
\be
m_{H_\pm}^2=\tm_2^2+\dfrac{1}{2}\tgl_3 v^2\,,
\ee
while a mixing is present in the neutral sector and the corresponding mass eigenvalues can be identified with the SM Higgs boson $h$, a heavier CP-even Higgs $H$, and a heavier CP-odd Higgs $A$.  At  the second order in the expansion $v^2/\tm_2^2$ their  masses read
\begin{align}
m_{h}^2&= v^2\left\{ \tgl_1-\left|\tgl_6\right|^2 \dfrac{v^2}{\tm_2^2}-\dfrac{1}{2}\slb\left(2\tgl_1-\tgl_3-\tgl_4\right)\left|\tgl_6\right|^2-\dfrac{1}{2}\left(\tgl_5^\ast \tgl_6^2+\hc\right) \srb \left(\dfrac{v^2}{\tm_2^2}\right)^2\right\}\,,\nn\\
m_{A,H}^2&=\tm_2^2\left\{1+\dfrac{1}{2}\rlb \tgl_3+\tgl_4\mp \left|\tgl_5\right|\rrb \dfrac{v^2}{\tm_2^2}+\dfrac{1}{2}\left|\tgl_6\right|^2\Big( 1\mp \cos{(2\theta_1)}\Big)\left(\dfrac{v^2}{\tm_2^2}\right)^2 \right\}\,.
\label{HiggsMasses}
\end{align}
Notice that the mass splitting between the two heavy neutral states is determined by $\left|\tgl_5\right|$: 
\be
m_H^2-m_A^2=\left(\left|\tgl_5\right|+\left|\tgl_6\right|^2\cos(2\theta_1)\dfrac{v^2}{\tm_2^2}\right)v^2\,,
\ee
neglecting higher order terms in $v^2/\tm_2^2$. Notice that in the near the decoupling limit defined as~\cite{Haber:1989xc,Haber:1994mt}
\be
\tm_2^2\gg\tgl_iv^2\,,
\label{DecouplingLimit}
\ee
all the heavy states become degenerate with the mass converging to $\tm_2$. Non-degenerate effects have been studied at length in the literature, see for example Refs.~\cite{Dorsch:2016tab,Kling:2016opi}.

Special attention has been devoted in the literature to the mixing between the two CP-even scalars, customary defined as $\cos(\alpha-\beta)$, being $\alpha$ a rotation angle. This is refereed to as the alignment parameter~\cite{Craig:2013hca}, such that when it vanishes, $\cos(\alpha-\beta)=0$, the $125\GeV$ Higgs corresponds to the direction of the VEV in the field space, that identifies the alignment limit. Expressing this mixing parameter in an expansion in $v^2/\tm_2^2$, we find 
\begin{equation}
\label{eq:mixing}
    \cos(\beta-\alpha)=-|\widetilde{\gl}_6|\dfrac{v^2}{\widetilde{m}_2^2}+\mathcal{O}\left(\dfrac{v^4}{\widetilde{m}_2^4}\right)\,.
\end{equation}\\

The SM fermion couplings to the Higgs doublets  $\Phi_1$ and  $\Phi_2$ are also controlled by their $Z_2$ charges.  Any tree-level FCNC contribution from the scalar sector is avoided if   each sector of fermions,  up, down, and leptons, couples to only one of the doublets.  Then the Yukawa couplings of all three neutral Higgs mass eigenstates remain aligned to the  fermion mass matrices and are simultaneously diagonalised.
There are  four possible $Z_2$ charge assignments to fermions that ensure such a solution to the FCNC problem in the 2HDMs.   They are  summarised as follows:
\begin{center}
\begin{tabular}{c||c|c||c|c|c|c|c}
\toprule
Model & $\Phi_1$  & $\Phi_2$  & $Q_L$ & $u_R$ & $d_R$ & $L_L$ &$e_R$ \\
\midrule
Type I          & $+$   & $-$ & $+$ & $-$ & $-$ & $+$ & $-$ \\
Type II         & $+$   & $-$ & $+$ & $-$ & $+$ & $+$ & $+$ \\
Type III (X)    & $+$   & $-$ & $+$ & $-$ & $-$ & $+$ & $+$ \\
Type IV (Y)     & $+$   & $-$ & $+$ & $-$ & $+$ & $+$ & $-$ \\
\bottomrule
\end{tabular}
\end{center}

Although the introduction of the $Z_2$ symmetry guarantees that, not only the lightest Higgs $h$, but also the heavier CP-even $H$ and CP-odd $A$ neutral scalars do not give rise to FCNC contributions,  there are NP contributions to processes with flavour changing mediated by the charged Higgses $H^\pm$.

As no other scalar  beyond the $125\GeV$ one has been discovered so far, we assume that the heavier Higgses have sufficiently large masses to escape direct detection at present colliders. Their existence, however, may be revealed by studying deviations from the SM predictions of various observables at low-energies. For this reason, in the next section, we will introduce the effective description of the $Z_2$-symmetric 2HD models near the decoupling limit. In this limit,   all the heavier scalars  will be taken as degenerate in mass: we will comment later on some effects associated with the departure from this condition.

%
%
\subsection{The low-energy effective description}

Integrating out the heavy scalar degrees of freedom leads to a low-energy effective description: near the decoupling limit, as defined in Eq.~\eqref{DecouplingLimit},  it has  already been presented in Ref.~\cite{Egana-Ugrinovic:2015vgy}. We will report here a few results relevant to our analysis. 

The effective Lagrangian involving the lightest Higgs $h$ can be written as the sum of several different parts, in addition to the canonically normalised kinetic terms, 
\be
\sL_h^\text{eff}=\sL^\text{eff}_{VV}+\sL^\text{eff}_{Y}-V^\text{eff}\,.
\ee
The mass terms and the couplings of the massive gauge bosons with $h$ can be written as 
\ba
\sL^\text{eff}_{VV} \supset& \,\,\dfrac{1}{2}Z_\mu Z^\mu \rlb m_Z^2 +g_{hZZ}\,v\,h+g_{h^2ZZ}\,h^2+\ldots\rrb+\\
&+W^+_\mu W^{-\mu} \rlb m_W^2 +g_{hWW}\,v\,h+g_{h^2 WW}\,h^2\ldots\rrb\,,
\ea
where the dots refer to interactions with a higher number of $h$. The $Z$ and $W$ masses, $m_Z^2$ and $m_W^2$, at tree level coincide with the SM expressions,
\be
m_W=\dfrac{g\,v}{2}\,,\qquad\qquad
m_Z=\dfrac{\sqrt{g^2+g^{\prime2}}\,v}{2}\,,
\ee
where $g$ and $g'$ are the gauge coupling constants for the $SU(2)_L$ and $U(1)_Y$ symmetries, respectively. The triple and quartic couplings turn out to be
\ba
\label{eq:kappaVV}
&g_{hVV}=\dfrac{2\,m^2_V}{v^2}\slb 1-\dfrac{1}{2}\left|\tgl_6\right|^2\left(\dfrac{v^2}{\tm_2^2}\right)^2\srb\equiv \dfrac{2\,m^2_V}{v^2}\kappa_V\,,\\
&g_{h^2VV}=\dfrac{m^2_V}{v^2}\slb 1-3\left|\tgl_6\right|^2\left(\dfrac{v^2}{\tm_2^2}\right)^2\srb\,,
\ea
where, we have introduced in the first equation the definition of $\kappa_V$ to match the notation used in the literature for the deviations from the SM predictions, that are proportional to $|\tgl_6|^2$ and only appear at the next-to-leading (NLO)
order of the expansion in $v^2/{\widetilde m}^2_2$.

Moving to the Higgs-fermion effective couplings, the dominant corrections arise at the leading order (LO) and are  also proportional  to $\tgl_6$. We adopt the  $\kappa$-formalism for the Yukawa interactions~\cite{Peskin:2013xra,Lepage:2014fla,Brod:2013cka},
\be
-\sL^\text{eff}_Y \supset M_f\ov{f}f+\dfrac{M_f}{v}\,h\,\left(\kappa_f\, \ov{f}f + \widetilde{\kappa}_f\, \ov{f}i\gamma_5f\right)+\ldots\,,
\ee
where the dots refer again to interactions with a higher number of $h$, and $M_f$ stand for the fermion mass matrix in the flavour basis (the flavour contractions are left implicit). We can distinguish the CP-even and the CP-odd contributions, $\kappa_f$ and $\widetilde{\kappa}_f$, respectively, as follows:
\ba
\kappa_u=&\kappa_d=\kappa_e=
1-\zeta_f\,\Re\left[\tgl_6^\ast\,e^{-i\xi/2}\right]\dfrac{v^2}{\tm_2^2}\equiv1-\zeta_f\left|\tgl_6\right|\cos{(\rho)}\dfrac{v^2}{\tm_2^2}\,,\\
\widetilde\kappa_u=&\widetilde\kappa_d=\widetilde\kappa_e=
-\zeta_f\,\Im\left[\tgl_6^\ast\,e^{-i\xi/2}\right]\dfrac{v^2}{\tm_2^2}\equiv-\zeta_f\,\left|\tgl_6\right|\sin{(\rho)}\dfrac{v^2}{\tm_2^2}\,,
\label{kappas}
\ea
where $\rho\equiv \arg\left[\tgl_6^\ast\,e^{-i\xi/2}\right]$, and the $\zeta_f$ parameters indicates the type of 2HDM under consideration,
\begin{center}
\begin{tabular}{c||c|c|c|c}
\toprule
 & Type I & Type II  & Type III (X) & Type IV (Y) \\
\midrule
$\zeta_u$ & $\cot\beta$    & $\cot\beta$ & $\cot\beta$   & $\cot\beta$ \\
$\zeta_d$   & $\cot\beta$    & $-\tan\beta$ & $\cot\beta$   & $-\tan\beta$ \\
$\zeta_e$   & $\cot\beta$    & $-\tan\beta$ & $-\tan\beta$   & $\cot\beta$ \\
\bottomrule
\end{tabular}
\end{center}
It is worth  pointing out that, within any of the 2HDM types, the NP corrections to the Yukawa couplings are, not only flavour blind, but also universal,  up to $\tan\beta$ dependent coefficients. This is a stronger result than that obtained in the MFV case~\cite{Alonso-Gonzalez:2021jsa,Alonso-Gonzalez:2021tpo} (or in any other flavour-based context), where the NP deviations are aligned with the SM Yukawas, but are different for up-type quarks, down-type quarks and charged leptons. This has a particular impact for CP violating observables, as we will discuss in the next section. 

Finally, the effective scalar potential can be written as
\be
V^\text{eff}= \dfrac{1}{2}m_h^2h^2+\dfrac{g_{h^3}}{3!}\,v\,h^3+\dfrac{g_{h^4}}{4!}h^4+\ldots\,,
\ee
where the dots refer again to interactions with a higher number of $h$. The $m_h^2$ is defined in Eq.~\eqref{HiggsMasses} and, up to the  second order in the expansion $v^2/\tm_2^2$, the cubic and quartic self-couplings read
\ba
\label{eq:higgs-self}
g_{h^3}&=\dfrac{3\, m_h^2}{v^2}-6\,\left|\tgl_6\right|^2\,\dfrac{v^2}{\tm_2^2}-\dfrac{1}{2}\left\{ \left[21\,\tgl_1 -12\left(\tgl_3+\tgl_4\right)\right]\left|\tgl_6\right|^2-\rlb 6\,\tgl_5\tgl_6^{\ast2} +\hc\rrb\right\}\left(\dfrac{v^2}{\tm_2^2}\right)^2\,,\\
g_{h^4}&=\dfrac{3\, m_h^2}{v^2}-36\,\left|\tgl_6\right|^2\,\dfrac{v^2}{\tm_2^2}-\left\{\left[105\,\tgl_1 -60\left(\tgl_3+\tgl_4\right)\right]\left|\tgl_6\right|^2-\rlb 30\,\tgl_5\tgl_6^{\ast2} +\hc\rrb\right\}\left(\dfrac{v^2}{\tm_2^2}\right)^2\,.
\ea
The dominant corrections arise at the LO and depend only on $|\tgl_6|^2$.

Besides the effective Lagrangian involving the lightest Higgs $h$, indirect signals of the presence of the heavier Higgses manifest  themselves in the four fermion operators. Only six operators of the SMEFT Lagrangian receive contributions in the 2HDM setup at mass dimension $d=6$: according to the notation in Ref.~\cite{Grzadkowski:2010es},  those are $Q^{(1)}_{qu}$, $Q^{(1)}_{qd}$, $Q^{(1)}_{\ell e}$, $Q_{\ell edq}$, $Q^{(1)}_{quqd}$ and $Q^{(1)}_{lequ}$. We can write these interactions after the EW symmetry breaking  as  the following effective Lagrangian operators:
\ba
\label{eq:smeft-operators}
\sL_{4f}^\text{eff}=&
c_{ijrs}^{qu(0)}\ov{u_{L_i}}u_{R_j}\ov{u_{R_r}}u_{L_s}+
c_{ijrs}^{qu(\pm)}\ov{d_{L_i}}u_{R_j}\ov{u_{R_r}}d_{L_s}+\\
&+c_{ijrs}^{qd(0)}\ov{d_{L_i}}d_{R_j}\ov{d_{R_r}}d_{L_s}+
c_{ijrs}^{qd(\pm)}\ov{u_{L_i}}d_{R_j}\ov{d_{R_r}}u_{L_s}+\\
&+c_{ijrs}^{\ell e(0)}\ov{e_{L_i}}e_{R_j}\ov{e_{R_r}}e_{L_s}+
c_{ijrs}^{\ell e(\pm)}\ov{\nu_{L_i}}e_{R_j}\ov{e_{R_r}}\nu_{L_s}+\\
&+\Bigg[c_{ijrs}^{\ell edq(0)}\ov{e_{L_i}}e_{R_j}\ov{d_{R_r}}d_{L_s}+
c_{ijrs}^{\ell edq(\pm)}\ov{\nu_{L_i}}e_{R_j}\ov{d_{R_r}}u_{L_s}+\\
&\hspace{7mm}+c_{ijrs}^{quqd(0)}\ov{u_{L_i}}u_{R_j}\ov{d_{L_r}}d_{R_s}+
c_{ijrs}^{quqd(\pm)}\ov{d_{L_i}}u_{R_j}\ov{u_{L_r}}d_{R_s}+\\
&\hspace{7mm}+c_{ijrs}^{\ell equ(0)}\ov{e_{L_i}}e_{R_j}\ov{u_{L_r}}u_{R_s}+
c_{ijrs}^{\ell equ(\pm)}\ov{\nu_{L_i}}e_{R_j}\ov{d_{L_r}}u_{R_s}+\hc\Bigg]\,,
\ea
where the associated dimensional Wilson coefficients read 
\ba
\label{eq:smeft-Wilson}
c_{ijrs}^{qu(0)}&=\delta_{ij}\delta_{rs}\dfrac{\widehat{M}_{u_j}\widehat{M}_{u_s}}{2v^2}\dfrac{\zeta_u^2}{\tm_2^2}\,,\qquad\qquad
c_{ijrs}^{qu(\pm)}=\dfrac{V^\ast_{ji} \widehat{M}_{u_j}\widehat{M}_{u_r} V_{rs}}{2v^2}\dfrac{\zeta_u^2}{\tm_2^2}\,,\\
c_{ijrs}^{qd(0)}&=\delta_{ij}\delta_{rs}\dfrac{\widehat{M}_{d_j}\widehat{M}_{d_s}}{2v^2}\dfrac{\zeta_d^2}{\tm_2^2}\,,\qquad\qquad
c_{ijrs}^{qd(\pm)}=\dfrac{V_{ij} \widehat{M}_{d_j}\widehat{M}_{d_r} V^\ast_{sr}}{2v^2}\dfrac{\zeta_d^2}{\tm_2^2}\,,\\
c_{ijrs}^{\ell e(0)}&=c_{ijrs}^{\ell e(\pm)}=\delta_{ij}\delta_{rs}\dfrac{\widehat{M}_{\ell_j}\widehat{M}_{\ell_s}}{2v^2}\dfrac{\zeta_\ell^2}{\tm_2^2}\,,\\
c_{ijrs}^{\ell edq(0)}&=2\delta_{ij}\delta_{rs}\dfrac{\widehat{M}_{\ell_j} \widehat{M}_{d_s}}{v^2}\dfrac{\zeta_\ell\,\zeta_d}{\tm_2^2}\,,\qquad\qquad
c_{ijrs}^{\ell edq(\pm)}=2\delta_{ij}\dfrac{\widehat{M}_{\ell_j} \widehat{M}_{d_r} V^\ast_{sr}}{v^2}\dfrac{\zeta_\ell\,\zeta_d}{\tm_2^2}\,,\\
c_{ijrs}^{quqd(0)}&=2\delta_{ij}\delta_{rs}\dfrac{\widehat{M}_{u_j} \widehat{M}_{d_s}}{v^2}\dfrac{\zeta_u\,\zeta_d}{\tm_2^2}\,,\qquad\qquad
c_{ijrs}^{quqd(\pm)}=-2\dfrac{V^\ast_{ji}\widehat{M}_{u_j} V_{rs}\widehat{M}_{d_s}}{v^2}\dfrac{\zeta_u\,\zeta_d}{\tm_2^2}\,,\\
c_{ijrs}^{\ell equ(0)}&=2\delta_{ij}\delta_{rs}\dfrac{\widehat{M}_{\ell_j} \widehat{M}_{u_s}}{v^2}\dfrac{\zeta_\ell\,\zeta_u}{\tm_2^2}\,,\qquad\qquad
c_{ijrs}^{\ell equ(\pm)}=-2\delta_{ij}\dfrac{\widehat{M}_{\ell_j} V^\ast_{sr}\widehat{M}_{u_s}}{v^2}\dfrac{\zeta_\ell\,\zeta_u}{\tm_2^2}\,,
\ea
in the fermion mass basis defined by
\be
M_f=V^\dag_f\,\widehat{M}_f\,U_f\,,
\ee
where $V_i$ and $U_i$ are unitary transformations over the left-handed (right-handed) fields. The matrices $\widehat{M}_f$ are the diagonal mass matrices and the CKM matrix is defined as $V=V_d^\dag V_u$. These results are obtained with massless active neutrinos, otherwise we should also consider the presence of the PMNS matrix. As can be seen from the previous expressions, there are no $\tgl_i$ parameters entering  those Wilson coefficients, so $\tm_2$ is the only NP parameter present beyond $\tan\beta$.\\

The presented above effective Lagrangian consists of dimension $d=4$ and $d=6$ operators: the light Higgs boson self-couplings and those with the gauge bosons are dimensionless and the NP corrections appear at LO and NLO, respectively, in the $v^2/{\widetilde m}^2_2$ expansion; on the other hand, the four-fermion operators contribute to both charged current processes and to the flavour conserving neutral current processes, with their dimensional Wilson coefficients being suppressed by $1/{\widetilde m}^2_2$.
As expected, in the limit of $\tm_2^2\to\infty$ or $\tgl_6\to0$, the SM form is recovered.  

One would be tempted to work in the ${\cal O}\left(v^2/{\widetilde m}^2_2\right)$ order of precision and use the effective Lagrangian given above at the tree level. However, as we discuss in the next section,  this is not sufficient for the analysis of the flavour physics constraints. In the language of the effective Lagrangian, the first corrections to the FCNC processes appear at the 1-loop level, with the loops generated by the four-fermion operators given in Eqs.~\eqref{eq:smeft-operators} and \eqref{eq:smeft-Wilson}.  Such loops would contribute e.g. to the meson-antimeson mixings and rare radiative meson decays, giving the corresponding 4-fermion operator with the Wilson coefficient suppressed by $1/{\widetilde m}^2_2$. In the analysis of the next section, those effects are effectively taken into account providing lower bounds on the parameter ${\widetilde m}_2$.

%
\section{Model predictions and numerical Analysis}

The low-energy effective Lagrangian introduced in  the previous section depends only on few parameters and, therefore, in the leading order in the expansion in the $1/{\widetilde m}^2_2$ many correlated deviations from the SM are predicted.  We summarise them  briefly.  

\begin{enumerate}
\item[a)] All corrections from 4-fermion operators to flavour-changing processes depend on the (approximately) degenerate mass of the heavy states.  In principle,  this is a powerful test of the models.  Unfortunately, neither any significant deviation from the SM predictions is observed in the precision flavour data nor a new scalar(s) discovered  to test those predictions.  However,  one can use the vast number of flavour data to put lower bounds on the new scalar masses as a function of $\tan\beta$, for each type of the 2HDM.

\item[b)] As  seen in Eq.~\eqref{kappas},  the real   and imaginary parts of the Yukawa couplings depend, in addition to the value of $\tan\beta,$  on only one effective parameter, that is 
\be
C=\left|{\tilde\lambda}_6\right|\cos({\rho})\frac{v^2}{{\widetilde m}^2_2}\,,\qquad\qquad
\widetilde{C}=\left|{\tilde\lambda}_6\right|\sin({\rho})\frac{v^2}{{\widetilde m}^2_2}\,,
\label{C}
\ee
where $C$ is for the real  and $\widetilde{C}$ for the imaginary parts.  So, the models predict strong same correlations between different $\kappa_f$'s and $\widetilde\kappa_f$'s.  Not only they are the same for all fermions in each sector but also correlated between sectors for a given value of $\tan\beta$. So, the Higgs decay rates to fermion pairs are correlated to each other, and also the  fermion loop contributions to the EDM are correlated.  It is also  worth remembering about the searches for CP violating effects  in the Higgs decays by analysing various CP sensitive asymmetries~\cite{CMS:2021sdq}.
The interesting point is that the correlations do not depend on the values of the parameters $C$ and $\widetilde{C}$, although the magnitude of the deviations in the individual $\kappa$'s of course does.

Trading the parameters $C$ and $\widetilde{C}$  for e.g.  $\kappa_u$ and  $\widetilde\kappa_u$, respectively,  for  the other two $\delta\kappa_f=\kappa_f -1$ and for $\widetilde\kappa_f$ one gets:
\be
\delta\kappa_f =\frac{\zeta_f}{\zeta_u}\delta\kappa_u\,,\qquad\qquad
\widetilde\kappa_f=\frac{\zeta_f}{\zeta_u}\widetilde\kappa_u\,.
\ee
Checking the Table below Eq.~\eqref{kappas},  we can see which are the most favorable Higgs boson decay modes to observe the deviations from the SM predicted by the models.   For instance,   for Type II,  such deviations in the down quark and lepton Yukawa couplings,  relevant for  the decays into a pair of the down  quarks  or leptons,  are enhanced by $\tan\beta^2$ (for $\tan\beta >1$) compared to the up quarks.  Similarly,  CP sensitive asymmetries in the $h\rightarrow\tau\tau$ are more promising to be observed than the CP violation effects in the $htt$ coupling.

Again, since no statistically  significant deviations in the Higgs decays from the SM predictions are observed in the data (and only upper bounds for the EDMs exist) we perform a fit to the data to get bounds on the parameters $C$ and $\widetilde{C}$. They allow to estimate the room left by the present  data for observing the deviations from the SM predicted by the 2HDMs in the Yukawa couplings.

\item[c)] It is interesting to  resolve the bounds on the parameters $C$ and $\widetilde{C}$ as  bounds on $\tgl_6 \cos({\rho})$ and  $\tgl_6 \sin({\rho})$ as a function of the mass scale of the heavy scalars.  First,  this is a check of consistency with near the decoupling limit approximation. Second,  one needs it for predicting the expectations for the triple Higgs coupling values.
\end{enumerate}

In the next subsections we present the results of our fits. They generically depend on the type of 2HDM considered and throughout the next text we will adopt the following colour-code for the different realisations of $Z_2$-symmetry:
\begin{center}
{\includegraphics[width=0.7
\textwidth]{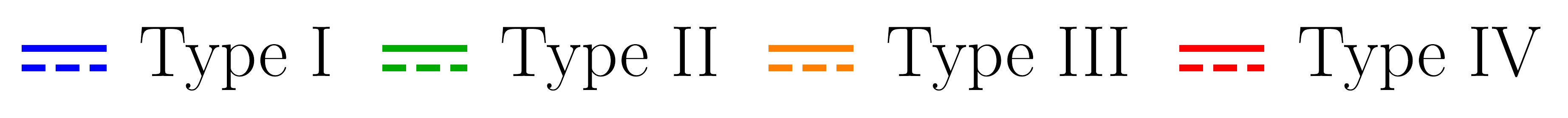}}
\end{center}

\subsection{Flavour Observables}

In this section, we discuss the lower bounds on the degenerate (near the decoupling limit) masses of the heavy scalars following the data for flavour observables. For the calculations of the SM predictions and their uncertainties, as well as those including NP, we use the software \textsc{Flavio}~\cite{Straub:2018kue}. 

The relevant four-fermion operators and the corresponding Wilson coefficients are listed in Eqs.~\eqref{eq:smeft-operators} and \eqref{eq:smeft-Wilson}, depending only on $\tm_2$ and $\tan\beta$ as NP parameters.  At tree level,   those operators contribute only to the charged current and flavour-conserving neutral current processes. 
In this category,  only  the $B^+\to\tau^+\nu$ decay provides an interesting constraint. 

Turning now to the very precisely measured FCNC processes,  as discussed in the previous section they receive contributions from the loops generated by the four fermion operators Eq.~\eqref{eq:smeft-operators}. Those loop effects cannot be neglected~\cite{Buras:2010mh} and $B_d$ and $B_s$ meson transitions are the most constraining observables of the $\tm_2$ \vs $\tan\beta$ parameter space.  We include them in an effective way 
by adopting the results presented in Refs.~\cite{Ciuchini:1997xe, Dai:1996vg, Logan:2000iv, Bobeth:2004jz, Enomoto:2015wbn, Cheng:2015yfu, Hu:2016gpe} for the one-loop contributions in the $Z_2$-symmetric 2HDM scenarios mediated by the five scalar bosons, and taking the mass degenerate limit for the heavier Higgses, to be consistent with  near the decoupling formalism. 

\begin{table}[tbh]
\centering
\resizebox{\textwidth}{!}{
\begin{tabular}{l||c|c|c}
\toprule
Observable & SM prediction & Exp. Measurement & Ref. \\
\midrule
$\mathcal{B}(B^0\to\mu^+\mu^-)$ & $1.1(1)\times 10^{-10}$ & $0.56(70)\times 10^{-10}$ & \cite{Altmannshofer:2021qrr}\\
$\ov{\mathcal{B}}(B_s\to\mu^+\mu^-)$ & $3.66(15)\times 10^{-9}$ & $2.93(35)\times 10^{-9}$ & \cite{Altmannshofer:2021qrr}\\
$\Delta M_{B_d}~[\text{GeV}]$ & $4.0(5)\times 10^{-13}$ & $3.334(13)\times 10^{-13}$ & \cite{HFLAV:2014fzu}\\
$\Delta M_{B_s}~[\text{GeV}]$ & $1.25(8)\times 10^{-11}$ & $1.1688(14)\times 10^{-11}$ & \cite{HFLAV:2014fzu}\\
$\mathcal{B}(B\to X_s \gamma)$ & $3.3(2)\times 10^{-4}$ & $3.27(14)\times 10^{-4}$ & \cite{Misiak:2017bgg}\\
$\mathcal{B}(B_s\to \phi\gamma)$ & $4.0(5)\times 10^{-5}$ & $(3.6 \pm 0.5 \pm 0.3 \pm 0.6)\times 10^{-5}$ & \cite{Belle:2014sac}\\
$\mathcal{B}(B^+\to\tau^+\nu)$ & $8.8(7) \times 10^{-5}$ & $1.09(24)\times 10^{-4}$ & \cite{ParticleDataGroup:2022pth}\\
$\mathcal{B}(B^0\to e^+e^-)$ & $2.7(3)\times 10^{-15}$ & $2.4(4.4)\times 10^{-9}$ & \cite{LHCb:2020pcv}\\
$\mathcal{B}(B_s\to e^+e^-)$ & $8.6(4)\times 10^{-14}$ & $0.30(1.29)\times 10^{-9}$ & \cite{LHCb:2020pcv}\\
$\mathcal{B}(B^0\to K^\ast\gamma)$ & $4.2(8)\times 10^{-5}$ & $4.33(15)\times 10^{-5}$ & \cite{HFLAV:2014fzu} \\
$\mathcal{B}(B^+\to K^\ast\gamma)$ & $4.2(9)\times 10^{-5}$ & $4.21(18)\times 10^{-5}$ & \cite{HFLAV:2014fzu} \\
$A_{CP}(B\to X \gamma)$ & $2(3)\times 10^{-18}$ & $0.032(34)$ & \cite{HFLAV:2014fzu} \\
$A_{CP}(B^0\to K^\ast \gamma)$ & $0.004(2)$ & $-0.002(15)$ & \cite{HFLAV:2014fzu} \\
$S_{\phi\gamma}$ & $-2(3)\times 10^{-4}$ & $0.43\pm0.30\pm0.11$ & \cite{LHCb:2019vks}\\
$C_{\phi\gamma}$ & $4(2)\times 10^{-3}$ & $0.11\pm 0.29\pm 0.11$ & \cite{LHCb:2019vks}\\
$\mathcal{A}^\Delta_{\phi\gamma}$ & $3(2)\times 10^{-2}$ & $-0.67^{+0.37}_{-0.41}\pm 0.17$ & \cite{LHCb:2019vks}\\
$\left<\dfrac{\diff\mathcal{B}'(B^+\to K^+\mu^+\mu^-)}{\diff q^2~\text{GeV}^{-2}}\right>_\text{bins}$ & - & - & \cite{LHCb:2014cxe}\\[4mm]
$\left<\dfrac{\diff\mathcal{B}'(B\to K^0{}^\ast\mu^+\mu^-)}{\diff q^2~\text{GeV}^{-2}}\right>_\text{bins}$ & - & -& \cite{LHCb:2016ykl}\\[4mm]
$[F_L, P_{1,2,3},P_{4,5,6,8}'](B\to K^0{}^\ast\mu^+\mu^-)$&-&-&\cite{LHCb:2020lmf}\\
\bottomrule
\end{tabular}
}
 \caption{\em Observables used in the fit. For the last three lines, data are collected in the corresponding references.
 }
\label{tab:observables-fit}
\end{table}

\begin{figure}[h!]
\centering
\subfigure[{}\label{fig:TI14}]
{\includegraphics[width=0.39\linewidth]{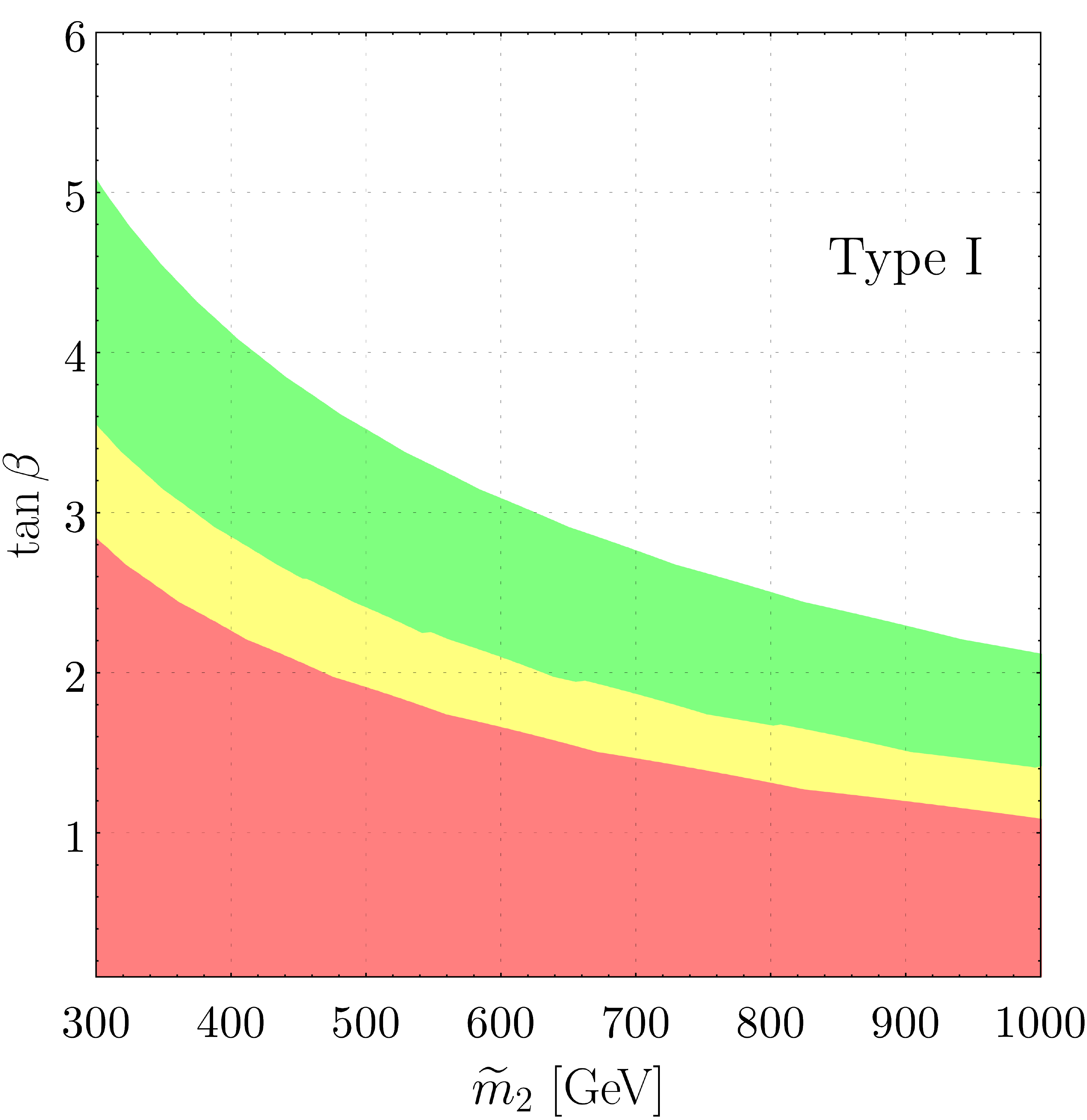}}
\subfigure[{}\label{fig:TII14}]
{\includegraphics[width=0.4
\linewidth]{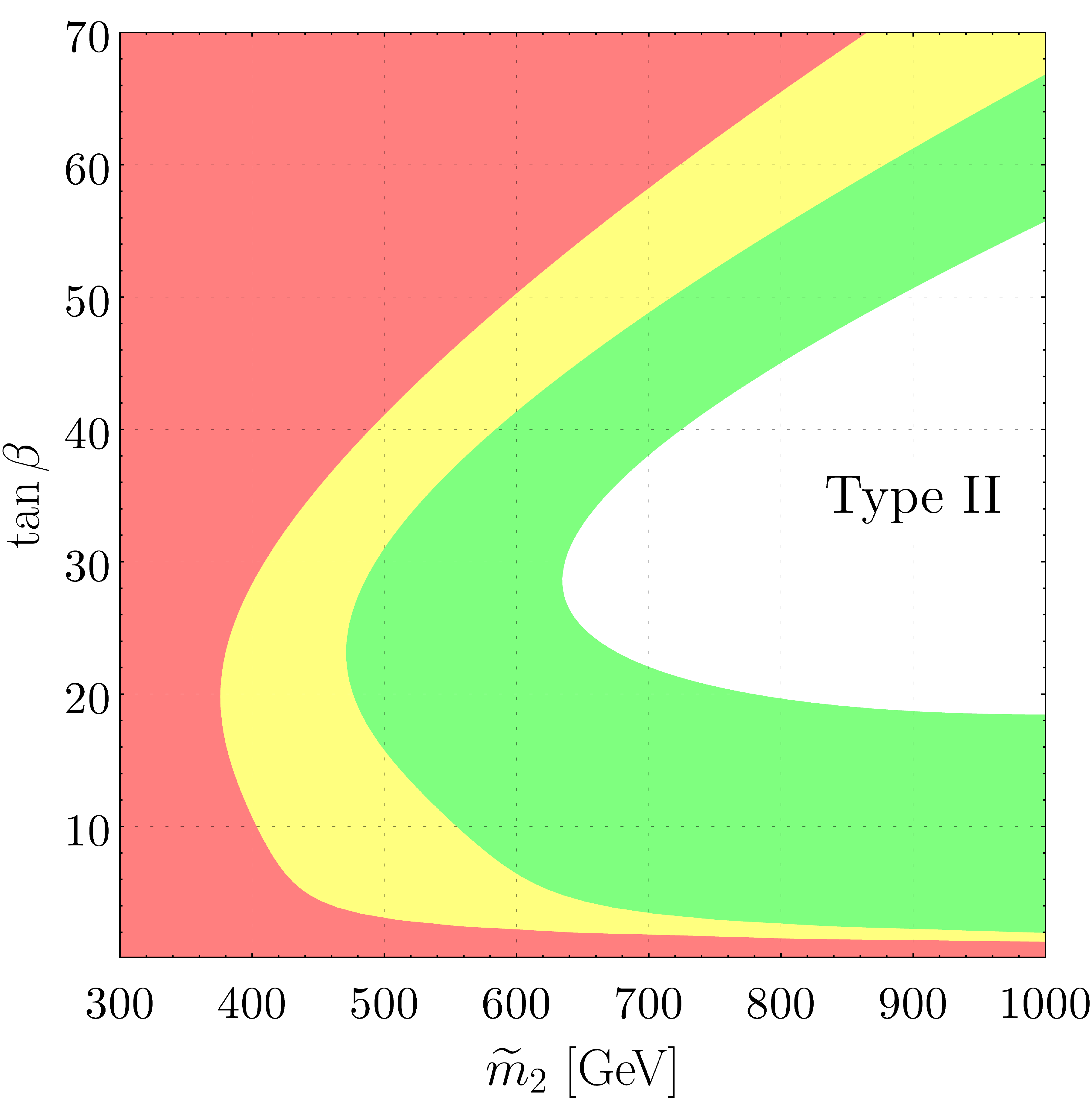}}
\subfigure[{}\label{fig:TIII14}]
{\includegraphics[width=0.4
\linewidth]{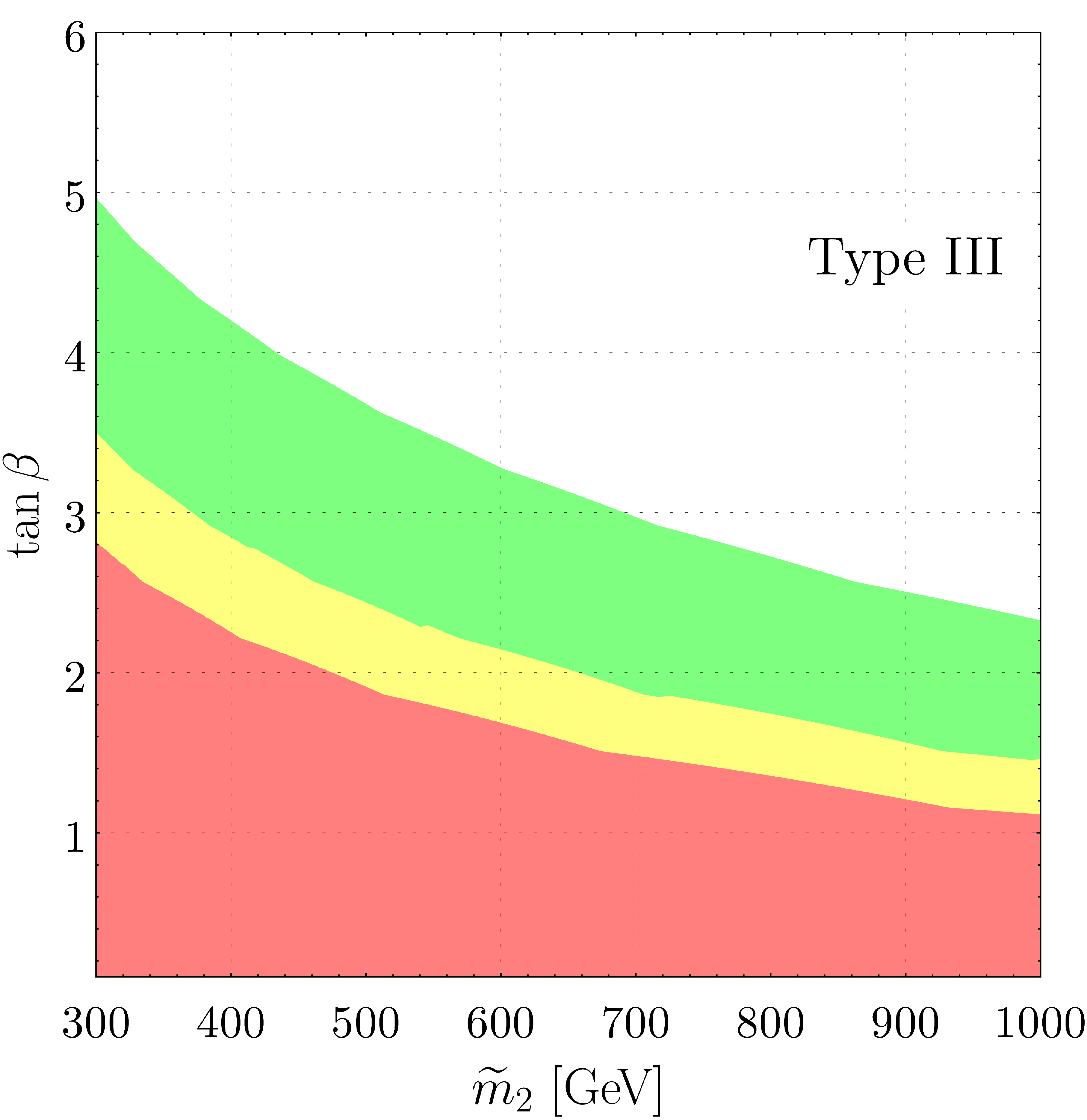}}
\subfigure[{}\label{fig:TIV14}]
{\includegraphics[width=0.39
\linewidth]{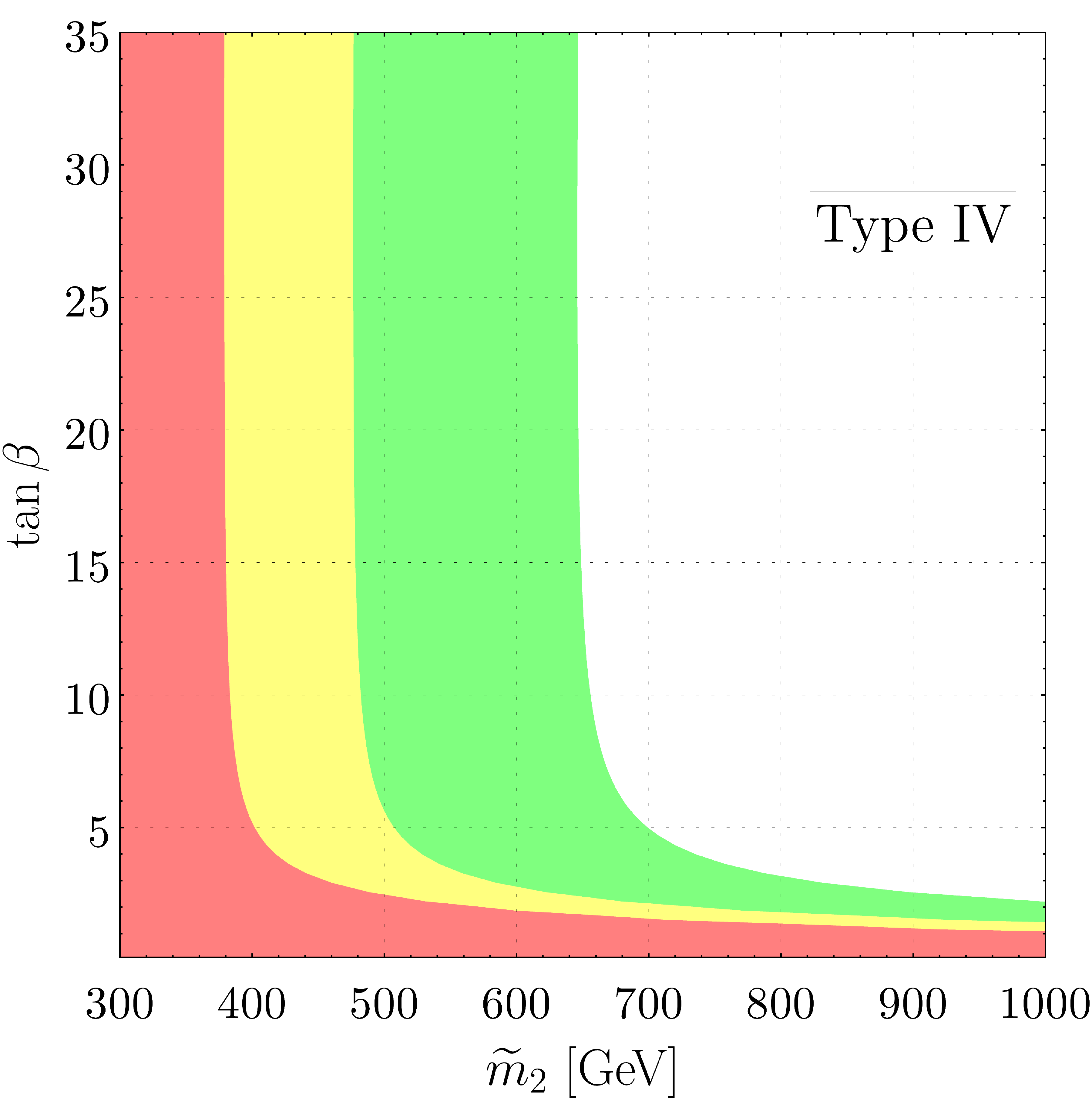}}
\par
\caption{\em Bounds given by flavour observables in Tab.~\ref{tab:observables-fit} for the different types of 2HDM. The green(yellow)[red] area corresponds to $[1\sigma,2\sigma]$($[2\sigma,3\sigma]$)[$>3\sigma$] region.}
\label{fig:flavour-14}
\end{figure}

The complete list of observables used in our fit is shown in Tab.~\ref{tab:observables-fit}. The plots referring to Type I and Type III are very similar to each other, excluding only the low values of $\tan\beta$, mainly due to $\mathcal{B}(B\to X_s\gamma)$, $\Delta M_{B_s}$ and $B_{d,s}\to\mu^+\mu^-$. The plot for Type IV shows, on top of the aforementioned limit for low $\tan\beta$ values, a strong cut on low $\tm_2$ for any values of $\tan\beta$, mainly constrained by the rare radiative $B$ decay. Finally, the most constrained parameter space is the one of Type II, where smallish and largish values of $\tan\beta$ are disfavoured even for large $\tm_2$. While $\mathcal{B}(B\to X_s\gamma)$ is the main responsible for the lower bound on $\tm_2$, the large $\tan\beta$ region is mostly constrained by $B_{s}\to\mu^+\mu^-$. For intermediate values of $\tan\beta$, the lower bound on $\tm_2$ is similar to the one for the Type IV scenario. This result follows mainly from the constraints given by the combination of the rates for the radiative $B$ decay and for the  leptonic neutral $B$ decays.

The reason for those similarities and differences may be traced back to the  $\tan\beta$ and $\cot\beta$ dependence of the Wilson coefficients  Eq.~\eqref{eq:smeft-Wilson}, which also determine the loop effects
 for the different types of 2HDM. In turn, they depend on the Yukawa couplings. As the Higgs couplings with the up-type quarks always include $\cot\beta$, they do not play a role in differentiating among the different constructions. On the other hand, $\zeta_d=-\tan\beta$ for Type II and Type IV and therefore $\mathcal{B}(B\to X_s \gamma)$ grows with larger $\tan\beta$ values: wherein the Type IV plot, this translates in the almost vertical bound on $\tm_2$ once $\tan\beta$ is larger than $\sim2$; in the Type II plot, instead, it corresponds to the bounds on $\tm_2$ for intermediate $\tan\beta$ values, that matches the limits aforementioned. This enhancement is absent in Type I and III as $\mathcal{B}(B\to X_s \gamma)$ is suppressed with $\tan\beta$. Finally, the differences between Type II and Type IV are mainly due to the Higgs couplings to the charged leptons that have an effect in the  leptonic neutral $B$ decays: their branching ratios are enhanced with $\tan\beta$ only for Type II and thus their bounds strongly constrain the parameter space.

The results presented above are overall in agreement with the literature, see e.g. Ref.s~\cite{Haller:2018nnx, Atkinson:2021eox, Atkinson:2022pcn, Beniwal:2022kyv}. One of the most updated references is Ref.~\cite{Haller:2018nnx}, which however presents two relevant differences. First, the lower bound on $\widetilde{m}_2$ present in Type II and IV is about $100\GeV$ smaller in our case. The reason is mostly to be found in the inclusion of recent data of angular observables of $B_0\to K^\ast \mu^+\mu^-$, as some of them show relatively large deviations from the SM predictions, even above $2\sigma$, e.g. $P_5'$.
Secondly, according to Refs.~\cite{Cheng:2015yfu,Haller:2018nnx}, in Type II the uppermost allowed value is $\tan\beta\sim30$, which is a bound almost twice as strong as ours. Such bound is mainly due to $B_s\to\mu^+\mu^-$ and $B^+\to\tau^+\nu$. The cause of the discrepancy could be one of the following reasons or a combination of them: i) a different treatment of the theoretical errors; ii) the fact that we do not let the masses of the different scalars float when doing the fit. Finally, our results for Type II are in excellent agreement with those of Ref.~\cite{Beniwal:2022kyv}.

Apart from flavour observables, one could expect significant constraints from direct searches at collider. The most constraining ones regard the searches for heavy neutral particles. Such bounds typically are given only for specific types of $Z_2$-models, especially for Type II. They depend on other parameters of the model besides $\tm_2$ and $\tan\beta$, such as the mixing of the CP-even scalars, $\cos(\beta-\alpha)$,  Eq.~\eqref{eq:mixing}. Direct searches bounds can be relevant for $\lesssim 400\GeV$~\cite{ATLAS:2020tlo}, that is in a region of the parameter space where the near the decoupling limit starts to be problematic. For this reason, we will not report such bounds in the results of the following sections.

\subsection{Electron electric dipole moment}

The present experimental bound on the electric dipole moment (EDM) of the electron provided by the ACME II collaboration~\cite{ACME:2018yjb},
\be
d_e=\left(4.3\pm3.1\pm2.6\right)\times 10^{-30}\text{ e cm}
\ee
translated to an upper bound at $2\sigma$ 
\be
|d_e|<1.24\times 10^{-29}\text{ e cm}\,,
\ee
represents one of the most powerful observables to constrain the CP-violating phases in NP models.  Using the effective description in terms of the $\kappa$-formalism in Eq.~\eqref{kappas}, we can translate this limit into a bound on $\widetilde{\kappa}_f$, the imaginary parts  in the Yukawa couplings  accounting for the effects of the $d=6$ corrections. 

\begin{figure}[tbh]
\centering
\subfigure[{Type II}\label{fig:kappa-tilde-T2}]
{\includegraphics[width=0.4
\linewidth]{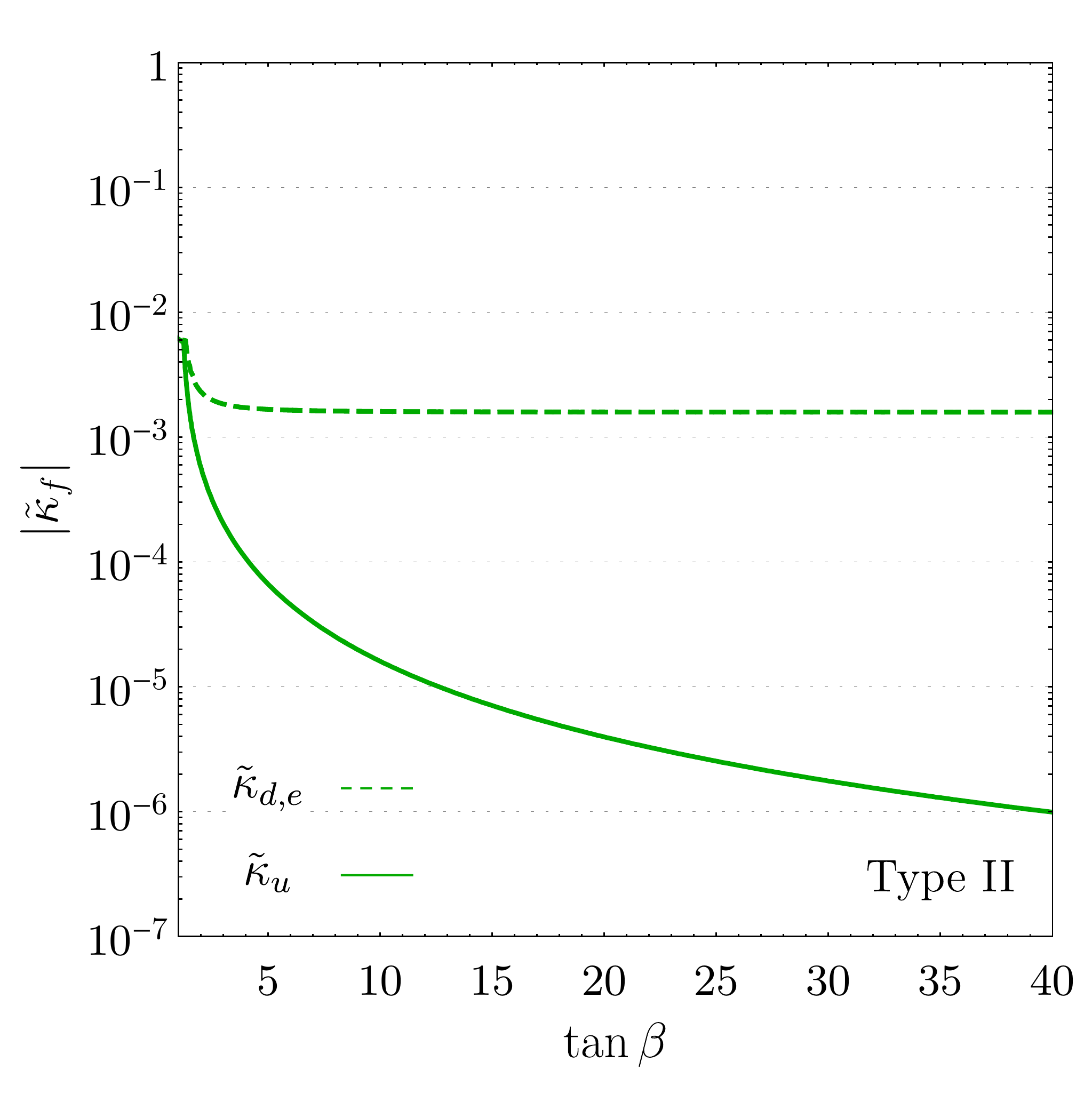}}
\subfigure[{Type IV}\label{fig:kappa-tilde-T4}]
{\includegraphics[width=0.4
\linewidth]{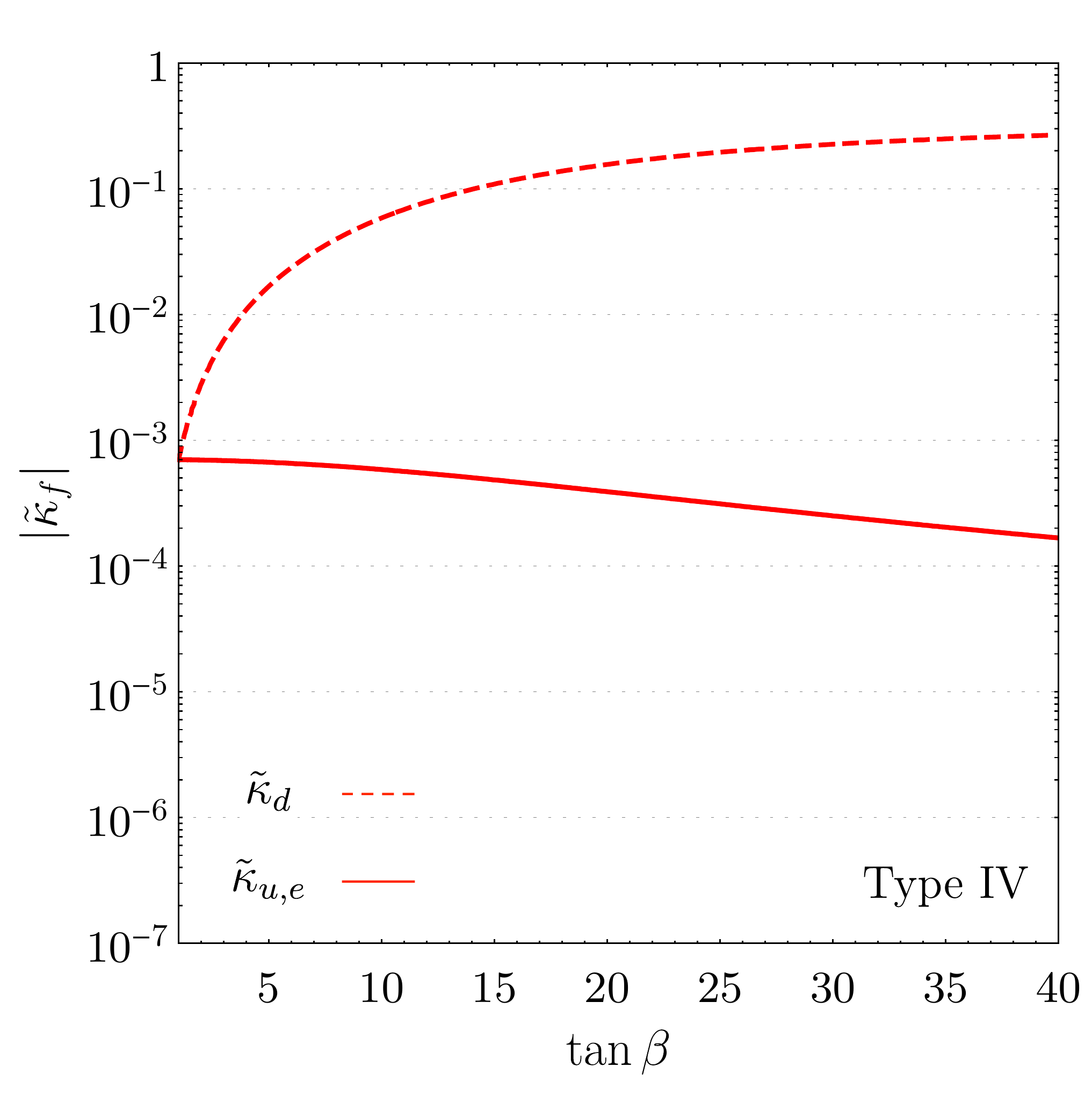}}
\par
\caption{\em Upper bounds on $\widetilde{\kappa}_f$ at 95\% C.L. for Type II and IV.}
\label{fig:kappa-tilde}
\end{figure}

Due to the correlations between $\widetilde\kappa_f$'s emphasized in the previous section, the electron EDM depends to a very good approximation only on the parameter $\widetilde{C}$ defined in Eq.~\eqref{C} and the value of $\tan\beta$, which fixes the relative magnitudes of the three $\widetilde\kappa_f$'s, for the up, down and charged lepton sectors.
Ref.~\cite{Altmannshofer:2020shb}, which corrects the results given in Ref.~\cite{Egana-Ugrinovic:2018fpy}, presents the explicit expression for $d_e$  in the limit we are discussing here, that is  in $Z_2$-symmetric 2HD models near the  decoupling limit:
\ba
\text{Type I:}\quad
&d_e\simeq 1.06\times 10^{-27}\,\left|\tgl_6\right|\sin{(\rho)}\left(\dfrac{1\TeV}{\tm_2}\right)^2\times\cot\beta\left[1+0.07\ln\left(\dfrac{\tm_2}{1\TeV}\right)\right]\text{e cm}\,,\\
\text{Type II:}\quad
&d_e\simeq -0.47\times 10^{-27}\,\left|\tgl_6\right|\sin{(\rho)}\left(\dfrac{1\TeV}{\tm_2}\right)^2\times\\
&\hspace{3cm}\times\left\{\tan\beta\left[1+0.16\ln\left(\dfrac{\tm_2}{1\TeV}\right)\right]-1.26\cot\beta\right\}\text{e cm}\,,\\
\text{Type III:}\quad
&d_e\simeq -0.47\times 10^{-27}\,\left|\tgl_6\right|\sin{(\rho)}\left(\dfrac{1\TeV}{\tm_2}\right)^2\times\\
&\hspace{3cm}\times\left\{\tan\beta\left[1+0.16\ln\left(\dfrac{\tm_2}{1\TeV}\right)\right]-1.25\cot\beta\right\}\text{e cm}\,,\\
\text{Type IV:}
\quad
&d_e\simeq 1.06\times 10^{-27}\,\left|\tgl_6\right|\sin{(\rho)}\left(\dfrac{1\TeV}{\tm_2}\right)^2\times\\
&\hspace{3cm}\times\left\{\cot\beta\left[1+0.07\ln\left(\dfrac{\tm_2}{1\TeV}\right)\right]+0.002\tan\beta\right\}\text{e cm}\,.
\ea
We use the above equations to find upper bounds of $\widetilde{C}$ as a function of $\tan\beta$ for each Type of the model: these limits are approximately independent from $\tm_2$, as the $\tm_2$-dependent logarithmic terms are numerically negligible in the considered range of $\tm_2$.  Then, using Eqs.~\eqref{kappas} one obtains upper bounds on the imaginary parts of the Yukawas shown in Fig.~\ref{fig:kappa-tilde}. They are shown for Type II and IV. For Type I the upper bounds for all 3 $\widetilde\kappa_f$'s are the same, of order $10^{-3}$; for Type III  the bound on $\widetilde\kappa_e$ remains as for Type II and the bound on $\widetilde\kappa_d$  is the same as the bound on $\widetilde\kappa_u$ in Type II.  This behaviour can be easily understood from Eqs.~\eqref{kappas}. 

It is worth commenting on the structure of the bounds in the context of experimental search for various CP-sensitive asymmetries in the Higgs production and decays.   At least in the framework of the $Z_2$ symmetric 2HDMs,  the search for such effects in the $htt$ coupling is least promising. Better chances are offered by the down quark sector in Type IV, but otherwise, the limits are always very strong,  at the level of $\widetilde\kappa_f\approx 10^{-2}-10^{-3}$.

The individual fermion loop contributions in the Barr-Zee diagrams to the electron EDM, and therefore the role of each loop in getting the final upper bound for $\tilde C$,  depend on the structure of the $\zeta_f$ coefficients in Eqs.~\eqref{kappas} and on the fermion masses, which enter into the loop calculation.  For the Type I case, for example,  all $\widetilde\kappa_f$'s  are weighted by  $\cot\beta$ and one can see that  the dominant contribution to $d_e$ comes from the top loop  (see the discussion in Ref.~\cite{Alonso-Gonzalez:2021jsa},  where the analysis considering only one NP parameter at a time was performed). Differently, for the Type II and Type III, the charged lepton contributions are enhanced by $\tan\beta$, and thus the bound can be directly associated with them. Finally, for type IV, it depends on the value of $\tan\beta$ that only enhances the down-type quarks.

It is also worth mentioning the possible cancellation that could take place for specific values of $\tan\beta$ between the terms in the curly brackets of the previous expressions. Interestingly, the same pattern of relations also holds for the neutron EDM, but with different numerical factors. As a result, even if one may select a specific value for $\tan\beta$ that cancels the NP contributions to the electron EDM, the bounds from the neutron EDM would still apply, providing a smaller but still strong constraint on the $\widetilde\kappa_f$.

\begin{figure}[tbh]
\centering
\subfigure[{}\label{fig:edm-3}]
{\includegraphics[width=0.4\linewidth]{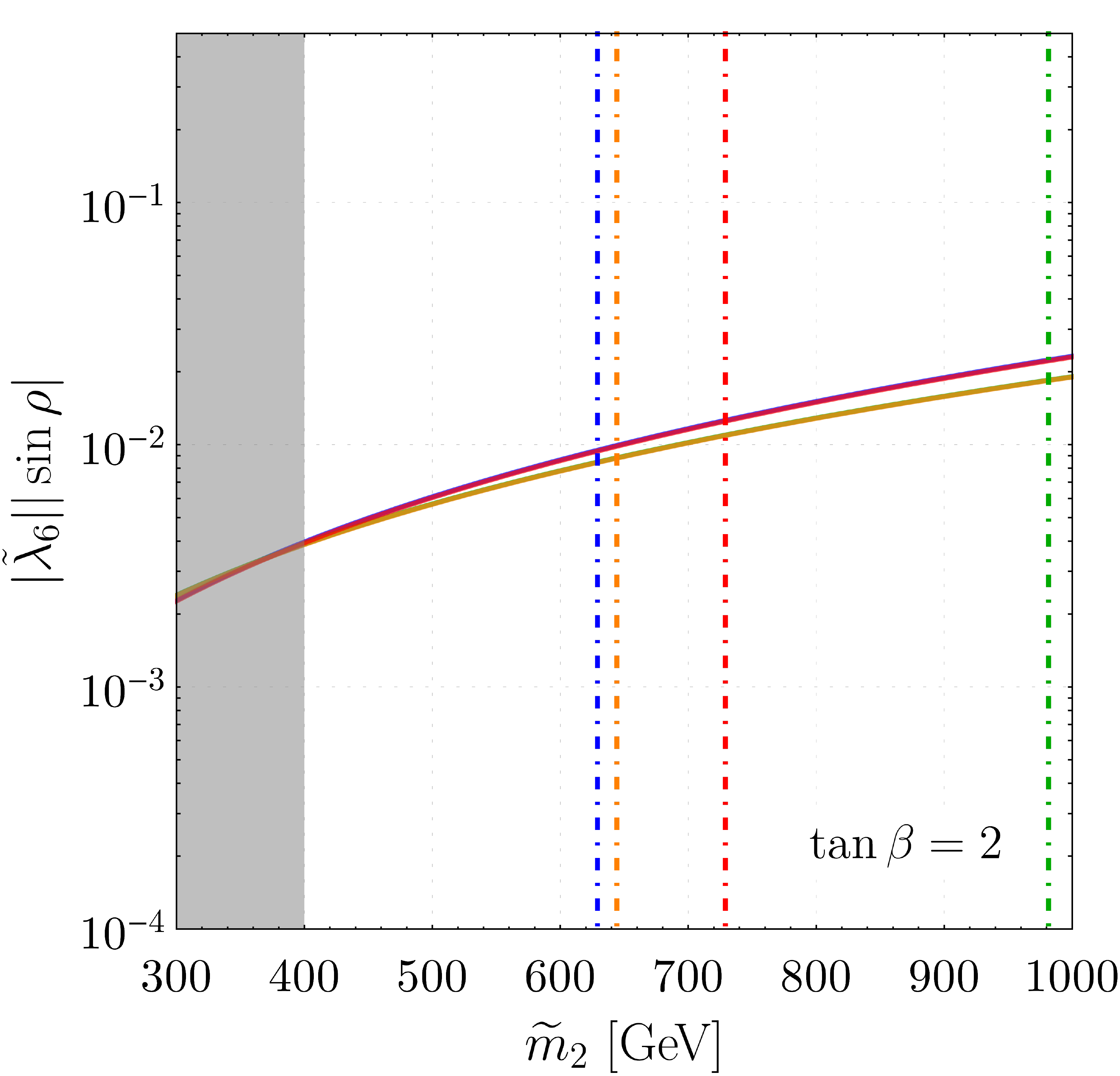}}
\subfigure[{}\label{fig:edm-15}]
{\includegraphics[width=0.4
\linewidth]{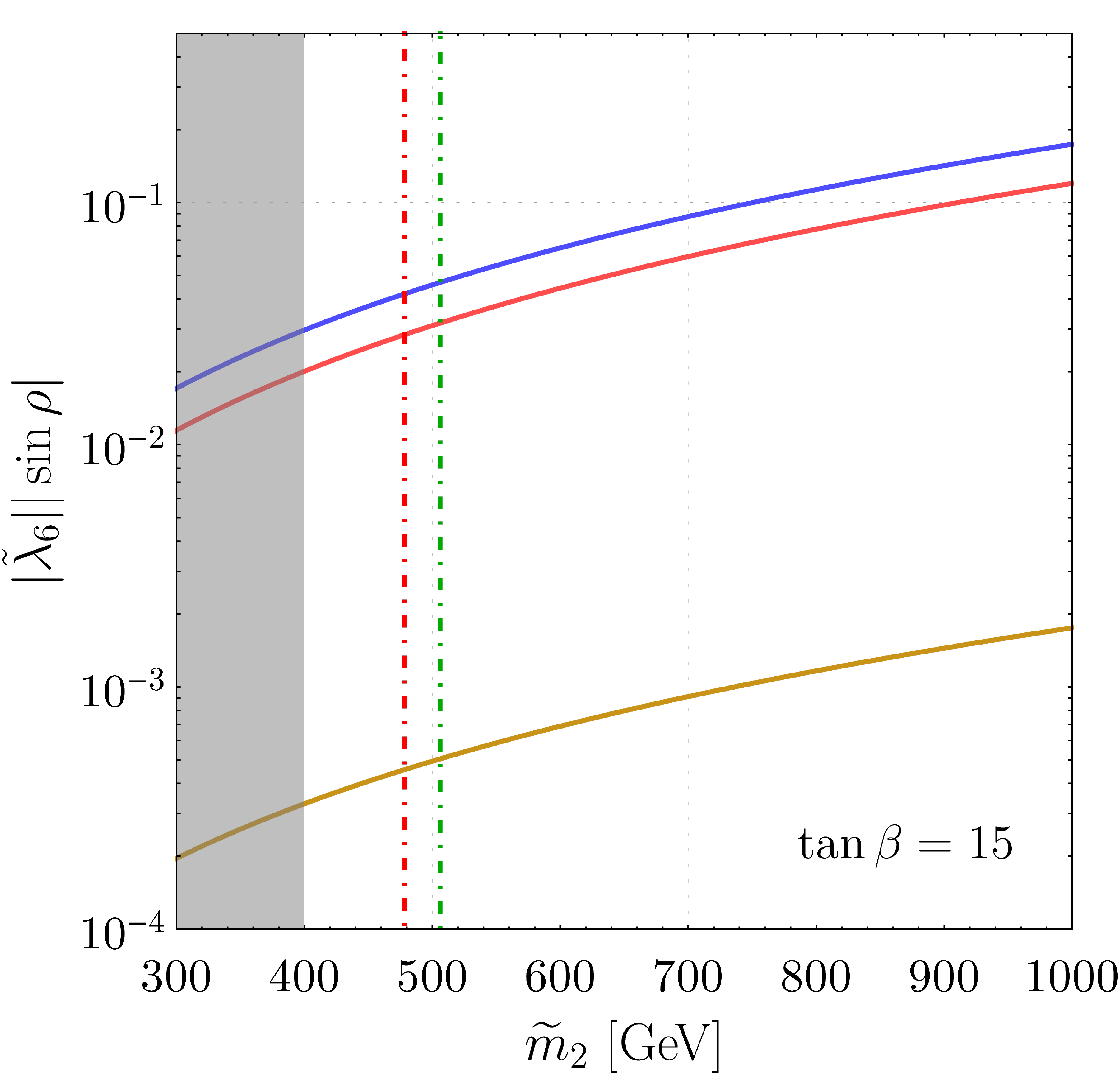}}
\par
\caption{\em Upper bounds at $2\sigma$ from electron EDM for different types of 2HDM and values of $\tan\beta$. Each colour corresponds to a specific realisation of 2HDM: Type I-blue, Type II-green, Type III-orange, Type IV-red. The plot is symmetric with respect to the x-axis for negative values of $\rho$. The area on the left-side of the vertical dot-dashed lines is excluded at $2\sigma$ by flavour observables. In the grey shaded area the $v/\widetilde{m}_2$ expansion is not precise and can be affected by $\gtrsim \mathcal{O}(40\%)$ corrections.}
\label{fig:edms}
\end{figure}

It is  useful  to show the upper bound on the parameter $\widetilde{C}$  as the  bound on the parameter $|\tgl_6|\sin{\rho}$ as a function of $\tm_2$ (the logarithmic terms are included). The results are shown in Fig.~\ref{fig:edms}. The four plots refers to different values of $\tan\beta=\{2,\,15\}$, and the lines are the $2\sigma$ upper bounds on $|\tgl_6|\sin\rho$ as a function of $\tm_2$. The colours identify the 2HDM construction: Type I-blue, Type II-green, Type III-orange, Type IV-red. The vertical lines are the $2\sigma$ lower bounds for $\tm_2$ as identified in Fig.~\ref{fig:flavour-14} from the fit with data from the flavour observables in Tab.~\ref{tab:observables-fit}. The values on the left of these lines are therefore excluded.

With analogous results for $|\tgl_6|\cos{\rho}$ as a function of $\tm_2$  obtained in the next subsection, one can infer  the acceptable values for the coupling $|\tgl_6|$ and the phase $\rho$, separately.

\subsection{Collider Observables}
In this section we investigate the bounds on the real parts of the Yukawa couplings following from the Higgs boson production and decay data. We employ the latest results from ATLAS~\cite{ATLAS:2022vkf} and CMS~\cite{CMS:2022dwd}. The experimental data can be used to constrain the parameters $\kappa_V$ describing the deviations from the SM predictions of the Higgs couplings to gauge bosons, defined in Eq.~\eqref{eq:kappaVV}, and the combinations
\be
r_f^2\equiv \kappa_f^2+\widetilde{\kappa}_f^2\,,
\ee
with $\kappa_f$ and $\widetilde{\kappa}_f$ in Eq.~\eqref{kappas}, for the couplings with fermions. 
The results reported by the experimental collaborations are shown in Tab.~\ref{tab:collider-data}.

\begin{table}[h!]
\centering
\begin{tabular}{l|c|c}
\toprule
& ATLAS~\cite{ATLAS:2022vkf} & CMS~\cite{CMS:2022dwd} \\
\midrule
$r_\mu$ &$1.07(26)$&$1.11(21)$\\
$r_\tau$ & $0.94(7)$ &$0.925(75)$\\
$r_b$ & $0.90(11)$ &$1.02(16)$\\
$r_t$ & $0.95(7)$ &$0.95(7)$\\
\midrule
$\kappa_W$ & $1.02(5)$ &$1.03(3)$\\
$\kappa_Z$ & $0.99(6)$ &$1.02(3)$\\
\bottomrule
\end{tabular}
 \caption{\em Values of $r_i$ and $k_V$ used for the fits.
}
\label{tab:collider-data}
\end{table}

\begin{figure}[tbh]
\centering
\subfigure[{}\label{fig:kappa-T2}]
{\includegraphics[width=0.4
\linewidth]{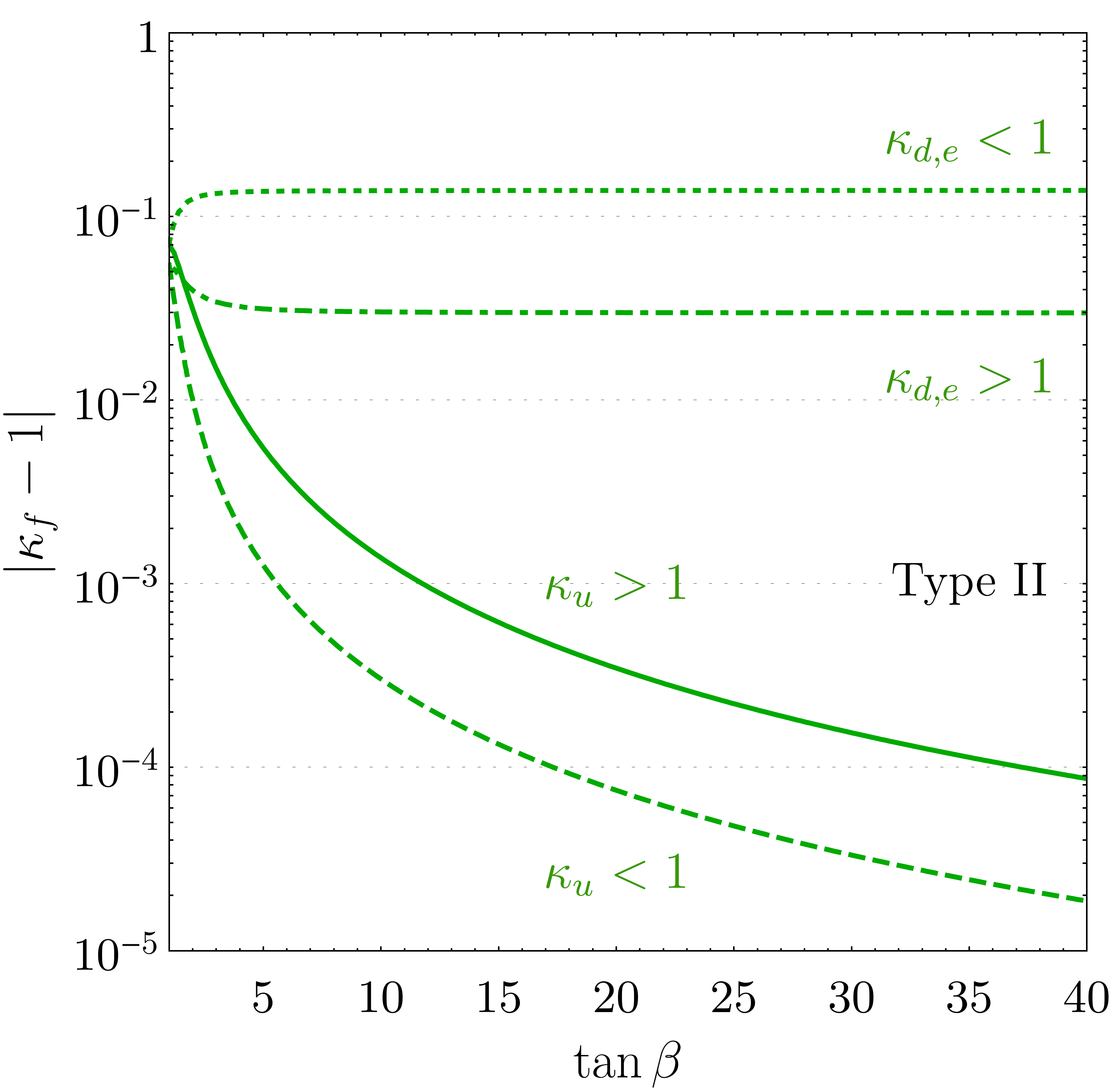}}
\subfigure[{}\label{fig:kappa-T4}]
{\includegraphics[width=0.4
\linewidth]{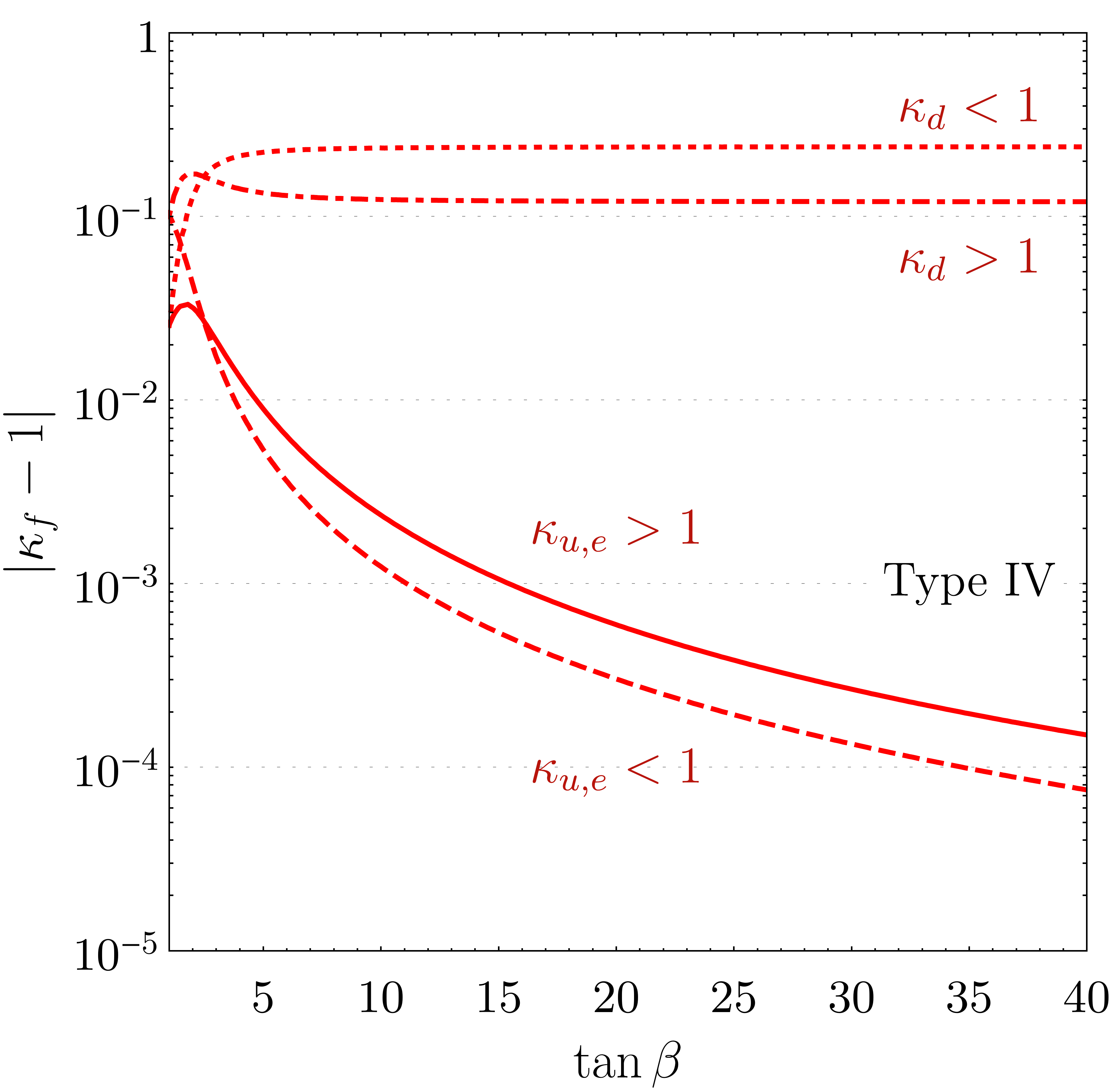}}
\par
\caption{\em Upper and lower bounds on $\kappa_f$ at 95\% C.L.  for Type II and IV.}
\label{fig:kappa}
\end{figure}

The EDM upper bounds at the $2\sigma$ level for the $\widetilde\kappa_f$'s are significantly stronger than the uncertainties on the $r_f$'s shown (at $1\sigma$) in Tab.~\ref{tab:collider-data},  so in our analysis of the bounds on the $\kappa_f$'s we assume $\widetilde\kappa_f=0$.  The accuracy of this approximation can be judged once the results for the bounds on  $\kappa$'s are obtained.   Thus,  taking $r_f=\kappa_f$,  we fit the parameter $C$ defined in Eq.~\eqref{C}. For each value of $\tan\beta$, we define a $\chi^2$-function based  on the data on $r_f$ and we obtain the best-fit value of $C$ for  each type of the model.  In each case, the results of the fit are most sensitive to those  $r_f$'s in Tab.~\ref{tab:collider-data} for which  $\zeta_f=\tan\beta$.  Next, using  Eqs.~\eqref{kappas} with fitted values for $C$, we obtain the bounds shown at $2\sigma$ level in  Fig.~\ref{fig:kappa} for Type II and IV. 
As it follows from Eqs.~\eqref{kappas},   for Type I the bounds are $\tan\beta$ independent and
the same for all $\kappa_f$'s in the range 
\be
0.9<\kappa_f<1.02\,,
\ee
For Type III,  $\kappa_d=\kappa_u$  and $\kappa_e$  remains as in Type II.  We see that in most cases $\kappa_f \gg\widetilde\kappa_f$, consistently with our approximation.
Similarly as for the EDM bounds, we note that the structure of correlations for the Yukawa couplings predicted by the models results in very strong bounds on the
$htt$ coupling.  For seeing the deviations from the SM, the Higgs decays to the down quarks and leptons (for Type II),  only to leptons (Type III) and only to the down quarks (Type IV) are much more promising.  

\begin{figure}[tbh]
\centering
\subfigure[{}\label{fig:col-3}]
{\includegraphics[width=0.4\linewidth]{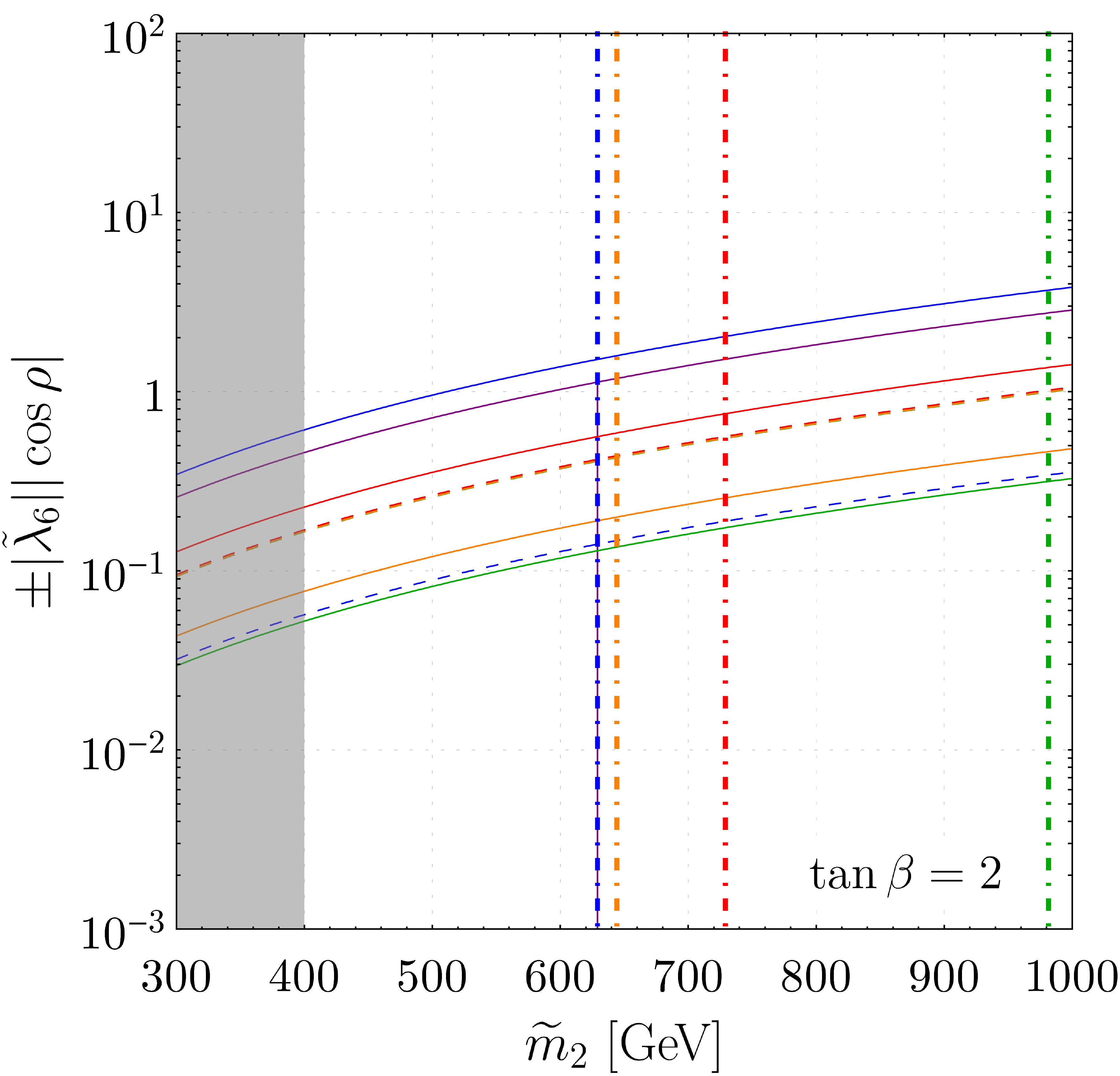}}
\subfigure[{}\label{fig:col-15}]
{\includegraphics[width=0.4
\linewidth]{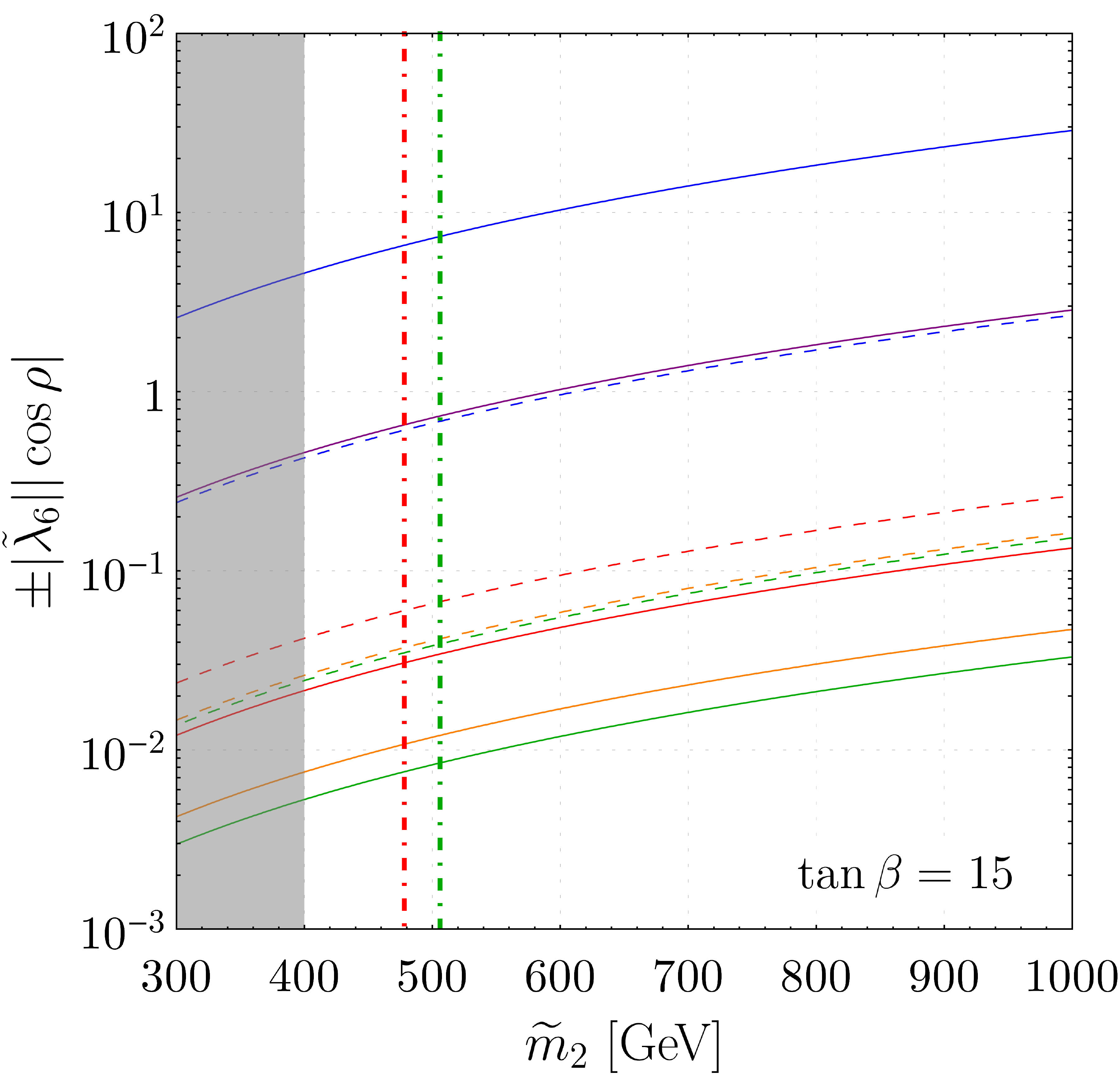}}
\par
\caption{\em Upper bounds at $2\sigma$ from collider observables for different types of 2HDM and values of $\tan\beta$. Each colour corresponds to a specific realisation of 2HDM: Type I-blue, Type II-green, Type III-orange, Type IV-red. The case with $\cos\rho>0~(<0)$ corresponds to solid (dashed) lines. The area on the left-side of the vertical dot-dashed lines is excluded at $2\sigma$ by flavour observables. The purple line corresponds to the limit in Eq.~\eqref{eq:fit-VV} and it applies to the case $\left|\cos\rho\right|=1$. In the grey shaded area the $v/\widetilde{m}_2$ expansion is not precise and can be affected by $\gtrsim \mathcal{O}(40\%)$ corrections.}
\label{fig:collider-14}
\end{figure}

The best fit results obtained for the parameters $C$ translate into parabola-like branches in the parameter space $|\tgl_6|\cos{(\rho)}$ \vs $\tm_2$. The results are shown in Fig.~\ref{fig:collider-14} for different values of $\tan\beta=\{2,\,15\}$.
The lines represent the $2\sigma$ upper limit on the combination $|\tgl_6|\left|\cos{(\rho)}\right|$: solid (dot-dashed) lines correspond to positive (negative) values of $\cos{(\rho)}$. As for Fig.~\ref{fig:edms}, the colours identify the 2HDM construction, that is Type I-blue, Type II-green, Type III-orange, Type IV-red. Moreover, the vertical lines are again the $2\sigma$ lower bounds for $\tm_2$ as identified in Fig.~\ref{fig:flavour-14}.
As the value of $\tan\beta$ grows, larger values for $|\tgl_6|\left|\cos{(\rho)}\right|$ are allowed in the Type I 2HDM case, while the contrary holds for the other cases. This is easy to understand for the Type I 2HDM, where $\zeta_{u,d,e}=\cot\beta$, and then the fitted values of the parameter $C$ increase with larger $\tan\beta$. On the other hand, the analysis is more involved for the other constructions: indeed, it is the combination of the sensitivity to the different $r_f$ and the dependence on $\tan\beta$ or $\cot\beta$ that determines which is the dominant contribution in the fit.

Comparing Figs.~\ref{fig:edms} and \ref{fig:collider-14}, we can conclude that generically $|\tgl_6|\left|\cos{(\rho)}\right|\gg \left|\tgl_6\right|\left|\sin{(\rho)}\right|$. This confirms once again the validity of the approximation chosen in the fit with only the fermionic collider data  and can be interpreted as $\left|\cos\rho\right|\approx1$.  
In fact, generically the bounds on $|\tgl_6|\left|\cos{(\rho)}\right|$ are $\mathcal{O}(10-100)$ stronger than the ones on $|\tgl_6|\left|\sin{(\rho)}\right|$.
The only exception lies in Type IV, as the two upper bounds become of the same order for $\tan\beta\sim 15$. Eventually, for larger values of $\tan\beta$, the bound on $|\tgl_6|\left|\sin{(\rho)}\right|$ become weaker than the one of $|\tgl_6|\left|\cos{(\rho)}\right|$.  One should thus expect corrections of $\mathcal{O}(1)$ which, however, should not change the magnitude of the limits.

On the other hand, some care is necessary when considering the large $\tan\beta$ scenario, as the bounds are relatively weak and $|\tgl_6|$ could reach values as large as $\tm_2^2/v^2$, that are inconsistent with the decoupling limit. This is already the case for $\tan\beta=15$ and for $\tm_2^2\gtrsim850\GeV$: in this case, the upper bound on $|\tgl_6|$ should be read from Eq.~\eqref{DecouplingLimit}.

Finally,  performing a  fit including only $\kappa_W$ and $\kappa_Z$ from both  collaborations, where
\ba
\kappa_V=&1-\dfrac{1}{2}\left|\tgl_6\right|^2\left(\dfrac{v^2}{\tm_2^2}\right)^2\,,
\ea
we obtain at $2\sigma$
\begin{equation}
\label{eq:fit-VV}
    \left|\tgl_6\right|\left(\dfrac{v^2}{\tm_2^2}\right)\leq 0.17\,.
\end{equation}
This result is valid for all four types of 2HDM realisations and is independent of $\tan\beta$. We show this limit in Fig.~\ref{fig:collider-14} as an additional purple line. Although strictly speaking, this line refers only to the case with $\cos\rho=1$, it helps to guide the eye in the interpretation of the relative magnitude of the different bounds.

\subsection{Triple Higgs Coupling}
With the results of the previous sections   for the bounds on the mass scale ${\tilde m}_2$ and the coupling  
${\widetilde\lambda}_6$, in this section we discuss the triple Higgs coupling, whose deviation from the SM is reported in Eq.~\eqref{eq:higgs-self}.   The  latest bounds  on it are given by the ATLAS collaboration~\cite{ATLAS:2022jtk} and read (at $2\sigma$)
\begin{equation}
\label{eq:triple-h-bound}
-1.4 < \dfrac{g_{h^3}}{g_{h^3}^\text{SM}} < 6.1\,.
\end{equation}

The leading NP contribution  depends on the ratio $|\tgl_6| ^2v^2/\tm_2^2$. It depends on  $\tan\beta$
through the values of 
${\widetilde\lambda}_6$. One then generically expects that, even considering the largest allowed values on $|\tgl_6|\left|\cos{(\rho)}\right|$ and $|\tgl_6|\left|\sin{(\rho)}\right|$ determined in the previous sections, the predicted deviation from the SM of $g_{h^3}$ is smaller than the direct bound in Eq.~\eqref{eq:triple-h-bound} (see for example Refs.~\cite{Arco:2020ucn,Arco:2022jrt,Arco:2022xum,Braathen:2019pxr,Braathen:2019zoh,Bahl:2022jnx}). 

It is well known that the radiative corrections to the triple Higgs coupling are relevant. Focussing on the $1-$loop contributions in the CP conserving limit, that is a good approximation given the smallness of the CP violating phases as shown in the previous sections, we can write~\cite{Kanemura:2002vm},
\be
g_{h^3}=g^{TL}_{h^3}+\delta g_{h^3}\,,
\ee
where $g^{TL}_{h^3}$ is given in Eq.~\eqref{eq:higgs-self} while $\delta g_{h^3}$ reads
\ba
\delta g_{h^3}\approx &\dfrac{3 m_{h}^2}{v^2}\left\{\dfrac{1}{12\pi^2 m_h^2 v^2}\slb m_{H}^4\rlb 1-\dfrac{\widetilde{m}_2^2}{m_{H}^2}\rrb^3+m_{A}^4\rlb 1-\dfrac{\widetilde{m}_2^2}{m_{A}^2}\rrb^3\srb\right.\\
&\hspace{1.5cm}\left.+\dfrac{ m_{H_\pm}^4}{6\pi^2 m_h^2 v^2}\rlb 1-\dfrac{\widetilde{m}_2^2}{m_{H_\pm}^2}\rrb^3-\dfrac{N_c m_t^4}{3\pi^2 m_h^2 v^2}\right\}\,,
\ea
where the last term is due to the SM contribution mediated by the top quark.

\begin{figure}[tbh]
    \centering
    \includegraphics[width=0.5\textwidth]{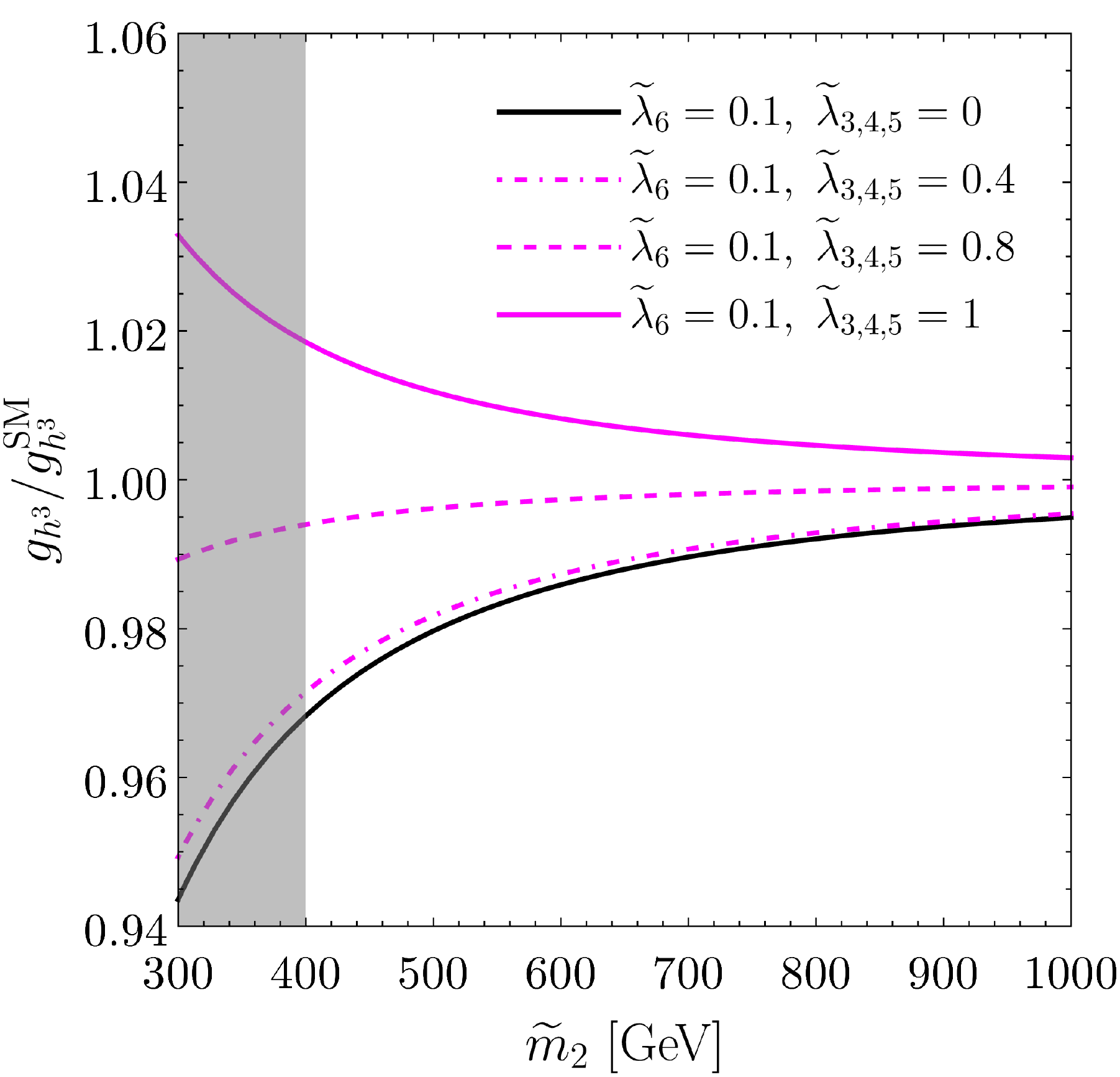}
    \caption{\em Relative importance of the loop- \textit{vs.} tree-level NP contributions. In the vertical axis we plot the ratio between the triple Higgs coupling prediction including NP contributions and the purely SM quantity at $1-$loop level. The solid-black line only contains the tree-level NP contributions, while the purple lines also contain the $1-$loop level NP contributions for different values of $\tgl_3=\tgl_4=\tgl_5$.}
    \label{fig:loop-tl}
\end{figure}

As can be seen, the contribution can be large if the decoupling is not realized or if the masses are not almost degenerate. In the case of almost decoupling, substituting the explicit expressions for the masses and keeping the leading contributions, the part containing the NP-corrections at $1$-loop simplifies and reads
\begin{equation}
   \delta g_{h^3}^\text{NP}\approx \dfrac{1}{16\pi^2}\left(\dfrac{v}{\widetilde{m}_2}\right)^2\left[(\tgl_3+\tgl_4+\tgl_5)^3+(\tgl^3-3\tgl_5(\tgl_3+\tgl_4)^2-\tgl_5^3)\right]\,.
\label{g3ratios}
\end{equation}
Notice that they are of the same order of magnitude as the leading corrections in Eq.~\eqref{eq:higgs-self} to the tree level result. Moreover, if $\tgl_i$ are positive, they may induce destructive interference with the tree-level NP contribution.
Notably, the dependence on $\tgl_{3,4,5}$ scales with the third power, so the relative importance between tree- and loop-level contributions highly depends on such couplings.
Such dependence is illustrated in Fig.~\ref{fig:loop-tl} for $\tgl_6=0.1$ and different values of $\tgl_{3,4,5}$. To maximize the effect, we take $\tgl_3=\tgl_4=\tgl_5$. 

One may expect that the parameters in the potential are of the same order of magnitude: for this reason and in order to simplify the discussion, without any loss of generality, in what follows we shall consider that $\tgl_{3,4,5}$ are close enough to $\tgl_6$ in order to neglect all NP contributions at $1$-loop.

\begin{figure}[h!]
\centering
\subfigure[{}\label{fig:col-TH-2}]
{\includegraphics[width=0.4\linewidth]{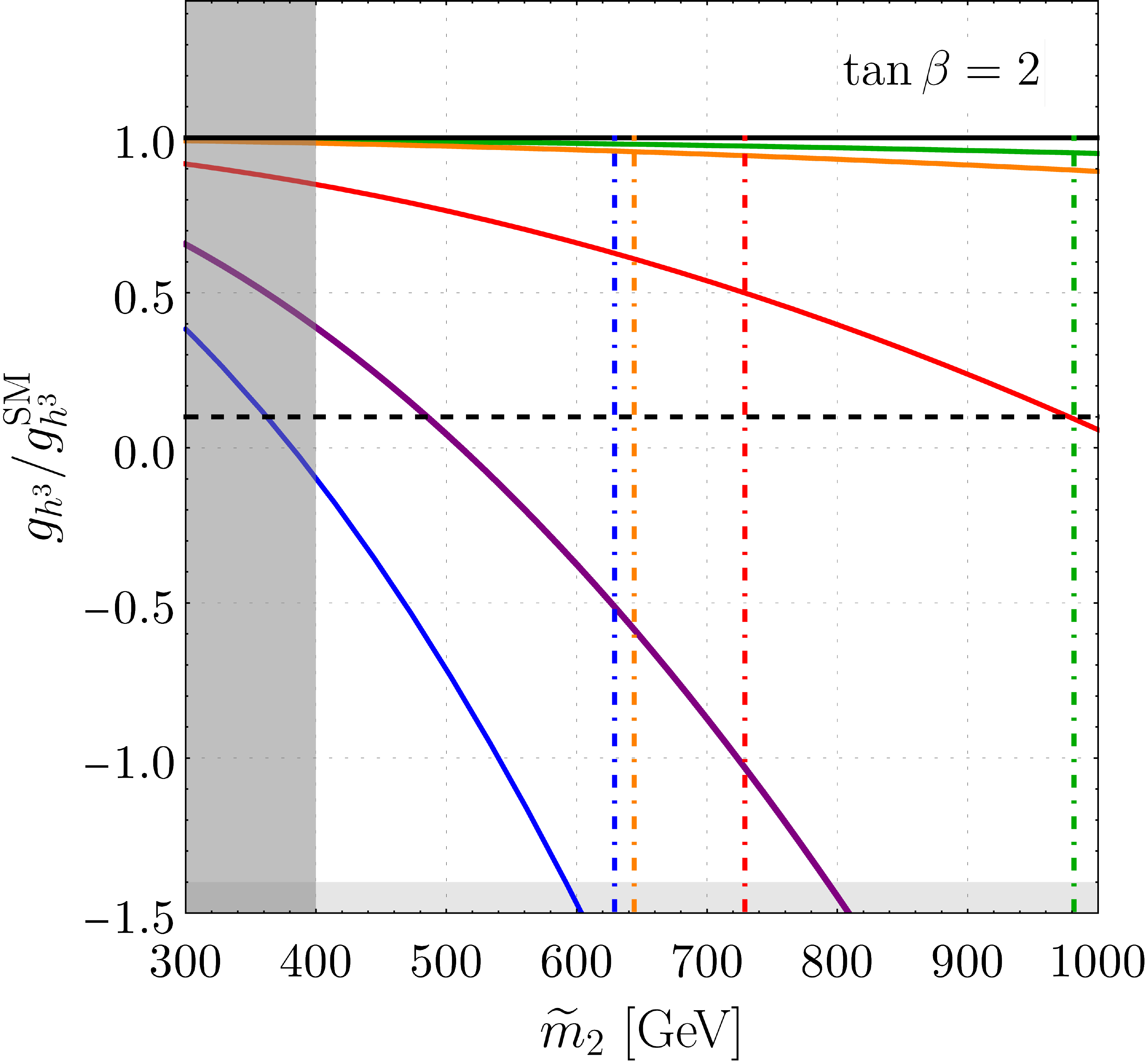}}
\subfigure[{}\label{fig:col-TH-15}]
{\includegraphics[width=0.4
\linewidth]{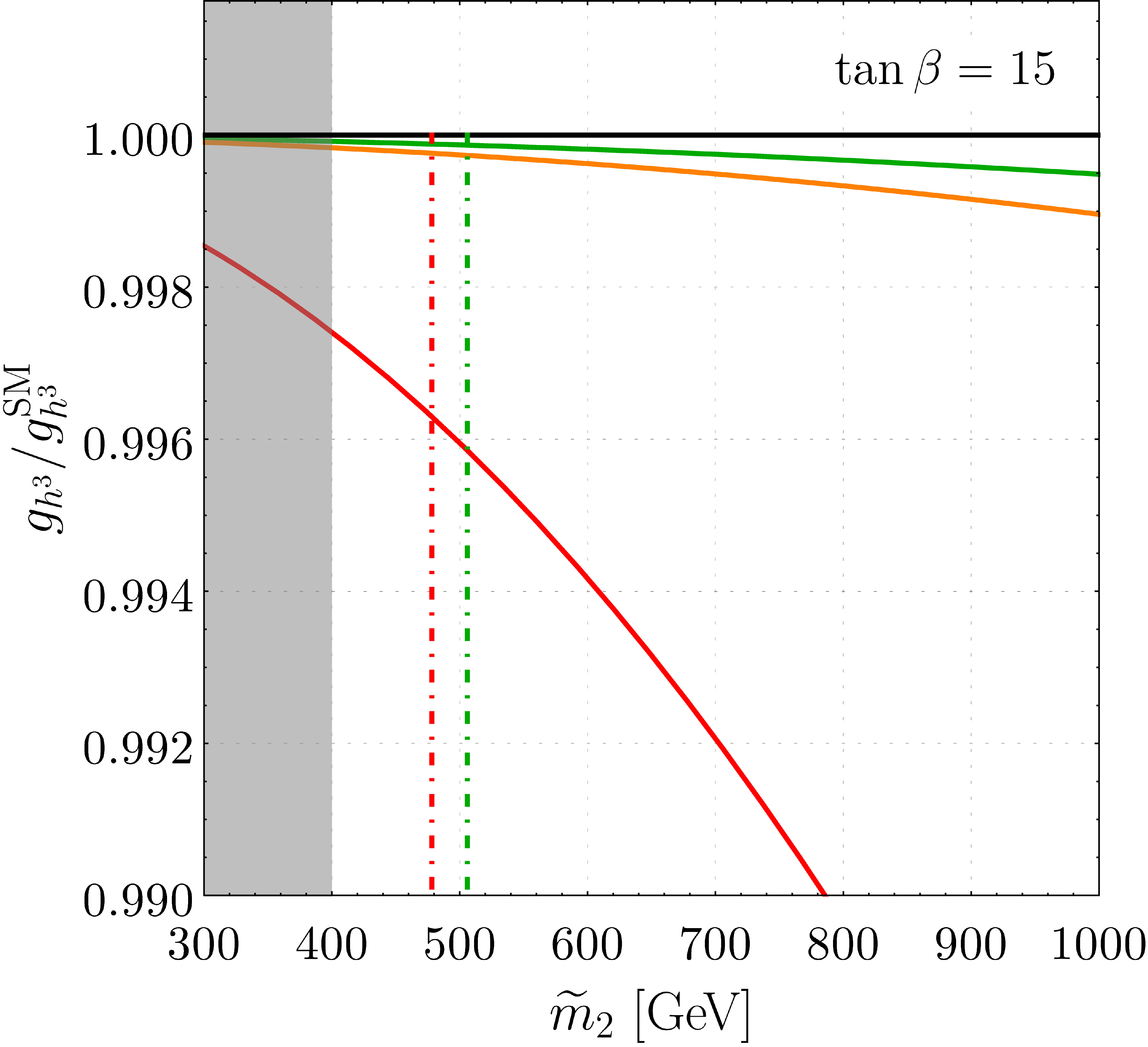}}
\par
\caption{\em Lower bounds on triple Higgs-coupling normalised to the SM (tree- and $1-$loop contribution) from collider observables for different types of 2HDM and values of $\tan\beta$. The solid black line corresponds to the SM prediction. Each colour corresponds to a specific realisation of 2HDM: Type I-blue, Type II-green, Type III-orange, Type IV-red. The purple line shows the universal bound from Higgs decay into gauge bosons in Eq.~\eqref{eq:fit-VV}. Only the area enclosed within the coloured and black solid lines is allowed. The shaded grey area is excluded at $2\sigma$ for all models from the direct measurement of the triple Higgs coupling. Finally, the area on the left-side of the vertical lines is excluded at $2\sigma$ by flavour observables. In the grey shaded area the $v/\widetilde{m}_2$ expansion is not precise and can be affected by $\gtrsim \mathcal{O}(40\%)$ corrections. The horizontal black dashed line corresponds to the lower value that HL-LHC will be able to probe at $95\%$C.L..}
\label{fig:collider-TH-14}
\end{figure}

We combine the information from the electron EDM and collider Higgs data of the previous sections to determine an upper bound on $|\tgl_6|^2 v^2/\tm_2^2$, summing quadratically the upper values for the real ($\propto\cos\rho$) and imaginary ($\propto\sin\rho$) parts, for any 2HDM realisations. With this procedure, we always obtain the strongest bound on the triple coupling as a function of ${\tilde m}_2$, except for the Type I 2HDM. In the latter case, the strongest bound coincides with Eq.~\eqref{eq:fit-VV} (for the values of $\tan\beta >1$). In Fig.~\ref{fig:collider-TH-14} we show  $2\sigma$ the lower bounds on $g_{h^3}/g_{h^3}^\text{SM}$, for  $\tan\beta=\{2,\,15\}$. We use the same colour code as the previous plots. The vertical lines are again the $2\sigma$ lower bounds for $\tm_2$ as identified in Fig.~\ref{fig:flavour-14}. Finally, the grey shaded area corresponds to the direct bound in Eq.~\eqref{eq:triple-h-bound}. When the purple and blue lines do not appear they are outside the scale of the plots (they give much weaker bounds).

For Type II and Type III, the models predict very small deviations from the SM for the triple Higgs couplings, for any values of $\tan\beta$. This is also true for Type IV for larger values of $\tan\beta$. On the other hand, the Type IV 2HDM for smaller $\tan\beta$ and the Type I for any $\tan\beta$ values sizeable deviations are allowed and can be probed in the future colliders.

Finally, HL-LHC prospects to measure much more precisely such quantity. The current projections estimate to test $g_{h^3}/g_{h^3}^\text{SM}\in[0.1,2.3]$ at $95\%$~C.L.~\cite{ATLAS:2022hsp}. This remarkable improvement will be very useful to constrain Type-I, but it will not restrict the parameter space for the other types with respect to the current bounds from Higgs-fermion interactions.

\boldmath
\subsection{$m_W$ from CDF II}
\unboldmath 

The determination of the W-boson mass, $m_W$, represents the last relevant anomaly in particle physics. The CDF II collaboration~\cite{CDF:2022hxs} measured it with the best-achieved sensitivity,
\be
m_W^\text{CDF II}=80.4335\pm0.0094\GeV\,,
\label{CFFIIMW}
\ee
showing a $7\sigma$ discrepancy relative to the SM prediction~\cite{Haller:2022eyb}
\be
m_W^\text{SM}=80.354\pm0.007\GeV\,.
\ee
Interestingly, this measure is not only in tension with the SM but also with all the other experimental determinations: the last one, from the ATLAS collaboration~\cite{ATLAS-CONF-2023-004}, shows no deviation from the SM prediction, although the uncertainties are not as good as those of the CDF II measurement. 

We refrain here from taking a definite point of view and instead discuss under which conditions the 2HDM setup we discussed so far can agree either with the SM prediction or with the experimental CDF II result\footnote{As reported in Ref.~\cite{Hays:2022qlw}, neglecting the possible correlations, one could estimate the average among the different experimental determinations obtaining a value for $m_w$ only $1\sigma$ far from the CDF II result, as expected. For this reason, we will perform our analysis considering the value in Eq. (1.1), understanding that the conclusions would remain invariant using instead the average quantity.}.

The most precise theoretical prediction of $m_W$ is obtained using $\alpha_\text{em}$, $G_F$, and $m_Z^2$ as experimental input. As the Fermi constant is extracted experimentally from the dominant muon decay, if NP contributes to the branching ratio of this decay, then $G_F$ gets redefined and a tree-level modification of $m_W$ follows. However, the impact of 2HDM the dominant muon decay induced by the operators of Eq.~\eqref{eq:smeft-operators} is extremely small, due to the suppression of the electron mass entering in the corresponding Wilson coefficients. It follows that no sizeable tree-level correction to the SM prediction is present. 

Another possibility concerns the role of oblique corrections, and more specifically of the Peskin-Takeuchi parameters, $S,T,U$~\cite{PhysRevD.46.381}.
 Denoted the loop-correction to the two-points function of the gauge fields $i\Pi^{\mu\nu}_{VV}\equiv i\Pi_{VV}\eta^{\mu\nu}+i\Pi^{pp}_{VV}p^\mu p^\nu$, they are defined as
\begin{align}
    &T\equiv \dfrac{1}{\alpha_\text{em}}\rlb \dfrac{\Pi^\text{NP}_{WW}(0)}{m_W^2}-\dfrac{\Pi^\text{NP}_{ZZ}(0)}{m_Z^2}\rrb\,,\\
    &S\equiv \dfrac{4c_W^2s_W^2}{\alpha_\text{em}}\slb \dfrac{\Pi^\text{NP}_{ZZ}(m_Z^2)-\Pi^\text{NP}_{ZZ}(0)}{m_Z^2}-\dfrac{c_W^2-s_W^2}{c_Ws_W}\dfrac{\Pi^\text{NP}_{Z\gamma}(m_Z^2)}{m_Z^2}-\dfrac{\Pi^\text{NP}_{\gamma\gamma}(m_Z^2)}{m_Z^2}\srb\,,\\
    &U\equiv \dfrac{4s_W^2}{\alpha_\text{em}}\slb \dfrac{\Pi^\text{NP}_{WW}(m_W^2)-\Pi^\text{NP}_{WW}(0)}{m_W^2}-\dfrac{c_W}{s_W}\dfrac{\Pi^\text{NP}_{Z\gamma}(m_Z^2)}{m_Z^2}-\dfrac{\Pi^\text{NP}_{\gamma\gamma}(m_Z^2)}{m_Z^2}\srb -S\,,
\end{align}
where $\alpha_\text{em}\equiv \alpha_\text{em}(m_Z)$ and the superscript ``NP'' indicates that the SM contribution has been subtracted and only NP contributes.
Their impact on $m_W$ reads~\cite{Grimus:2008nb}
\begin{equation}
    m_W^2=(m_{W}^\text{SM})^2\left(1+\dfrac{s_W^2}{c_W^2-s_W^2}\Delta r'\right)\,,
\end{equation}
where
\begin{equation}
    \Delta r'\equiv \dfrac{\alpha_\text{em}}{s_W^2}\left(-\dfrac{1}{2}S+c_W^2 T+\dfrac{c_W^2-s_W^2}{4s_W^2}U\right)\,.
\end{equation}

The expressions for the oblique parameters for multi-higgs extensions have already been computed in Refs.~\cite{Grimus:2007if, PhysRevD.83.055017, Funk:2011ad, Heo:2022dey,Bahl:2022xzi}.
In the alignment limit, these expressions can be expanded in terms of $v^2/\tm_2^2$ and, in the CP-conserving case, they read
\begin{align}
    &S\approx\dfrac{1}{24\pi}\tgl_4 \dfrac{v^2}{\tm_2^2}\,,\\
    &T\approx\dfrac{1}{192\pi^2 \alpha_\text{em}}(\tgl_4^2-|\tgl_5|^2)\dfrac{v^2}{\tm_2^2}\,.
\end{align}
The parameter $U$ has been neglected as it receives corrections at the next order in the expansion $v^2/\tm_2^2$. From the above expressions, we can estimate that $|T|\sim 8~|S|$, unless a cancellation occurs between the Lagrangian parameters $\tgl_4$ and $\tgl_5$. 

\begin{figure}[h!]
\centering
\subfigure[{}\label{fig:mw1}]
{\includegraphics[width=0.42\textwidth]{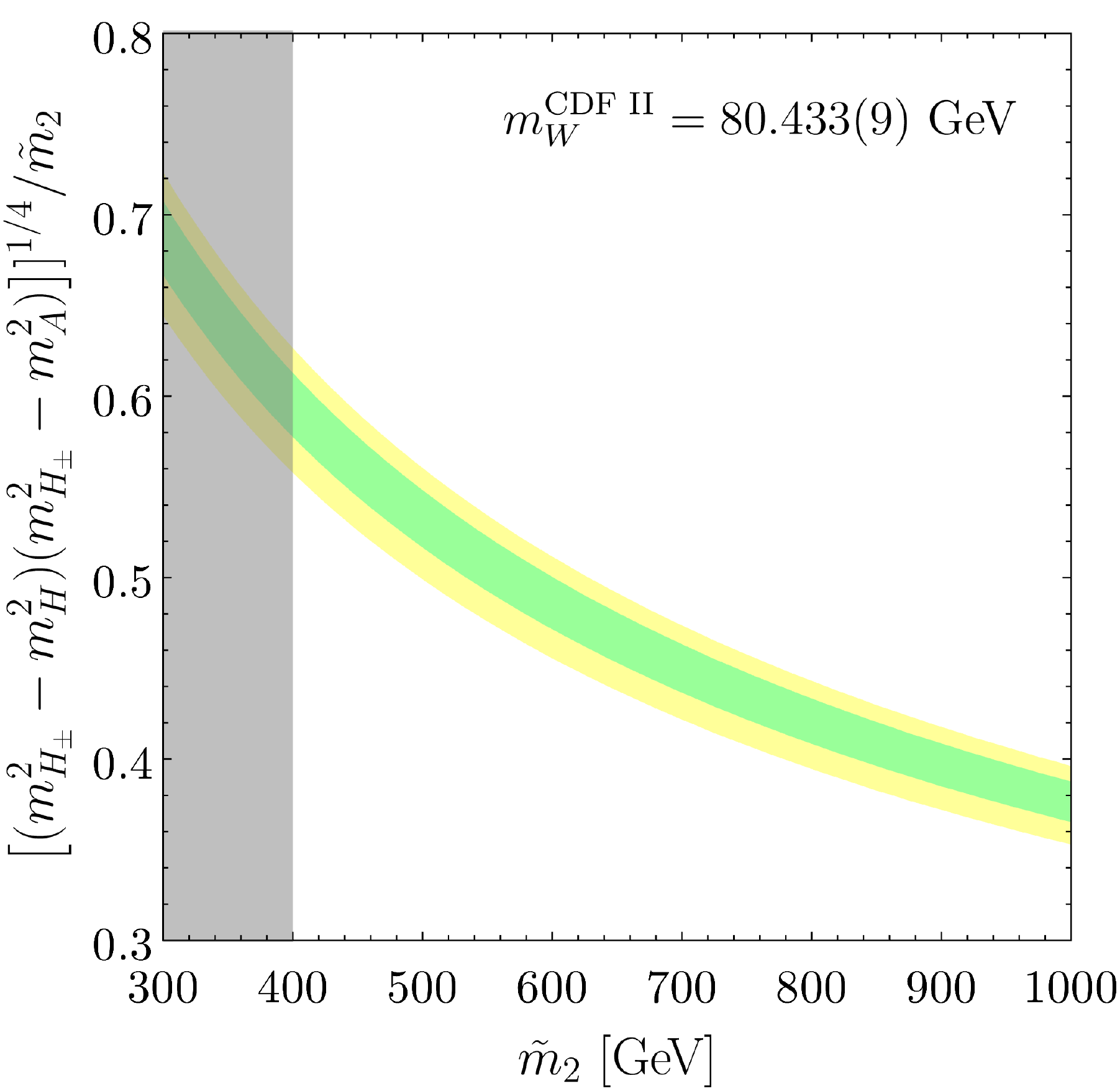}}
\subfigure[{}\label{fig:mw2}]
{\includegraphics[width=0.42
\textwidth]{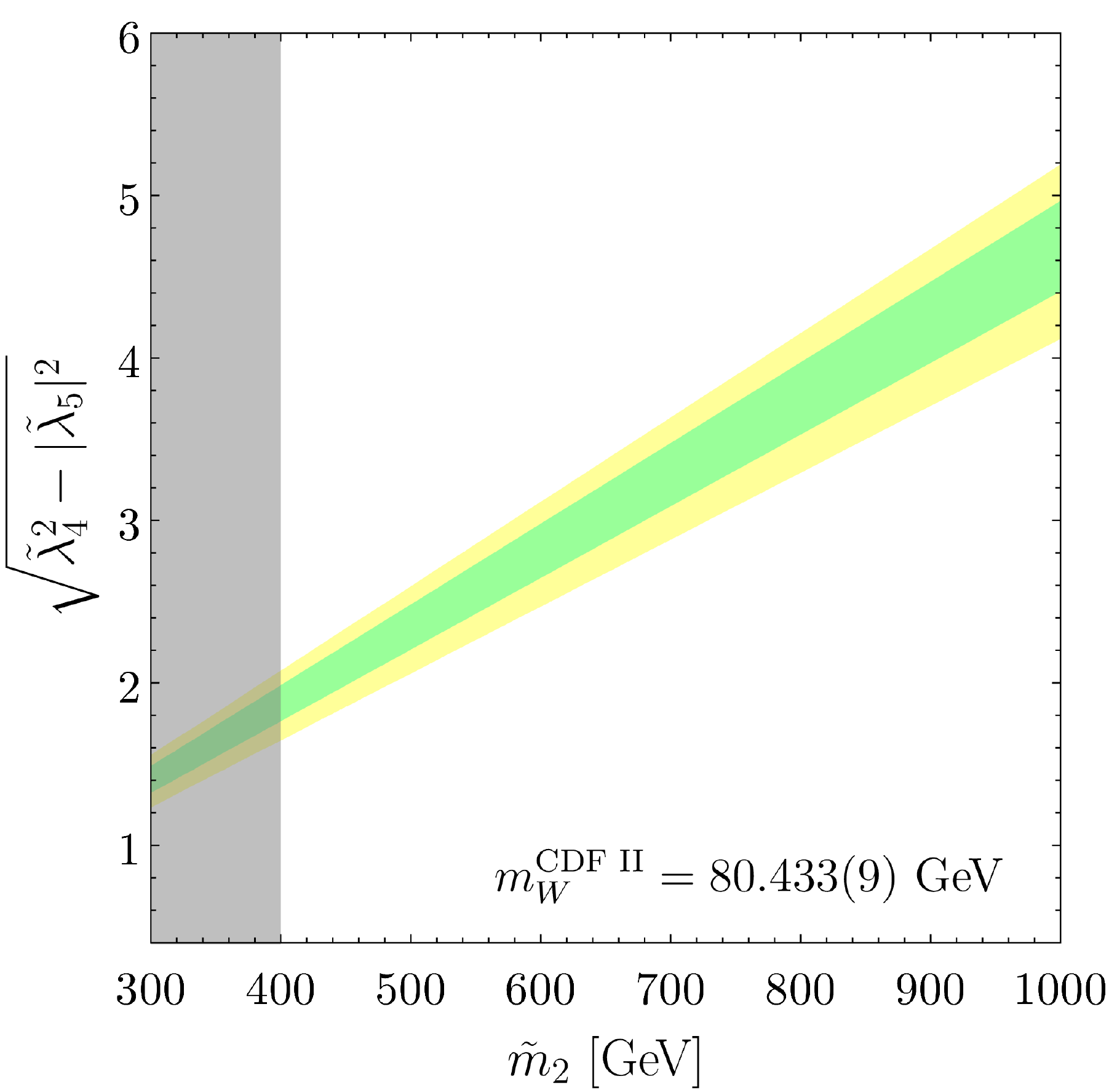}}
\par
\caption{\em Parameter space where it is possible to explain the CDF II measurement of $m_W$ via oblique $T$ parameter. In green(yellow) the experimental value at $1\sigma$($2\sigma$). In the grey shaded area the $v/\widetilde{m}_2$ expansion is not precise and can be affected by $\gtrsim \mathcal{O}(40\%)$ corrections.}
\label{fig:mw-plots}
\end{figure}

If one takes for true the CDF II anomaly and pretends to solve it within this context, the relative signs of the second expression are fixed, as $\tilde\lambda_4>|\tilde\lambda_5|$ guarantees a larger prediction than the SM one. 
The $T$ parameter is related to the violation of the custodial symmetry of the scalar potential and vanishes if all the heavy scalars are degenerate in mass. In fact, we can see that
\begin{equation}
    \tgl_4^2-|\tgl_5|^2\approx\dfrac{4}{v^4}(m_{H_\pm}^2-m_H^2)(m_{H_\pm}^2-m_A^2)\,.
\end{equation}
This implies that a minimal mass-splitting is necessary in order to explain the $m_W$ anomaly through a sizeable $T$ oblique parameter. Such result can be seen in Fig.~\ref{fig:mw-plots}.

It is interesting to notice that the explanation of $m_W$ fixes also the ordering of the mass hierarchy among the heavy scalars, leaving just two possibilities:
\begin{equation}
    m_{H_\pm}^2 > m_{H}^2,~m_{A}^2\,, \qquad \qquad \qquad m_{H_\pm}^2 < m_{H}^2,~m_{A}^2\,.
\end{equation}

On the other side, if the CDF II anomaly will lose evidence, this may be interpreted in the context of the $Z_2$-symmetric 2HDM as an indication of the (almost) degeneracy of the heavy scalar masses.

%
\section{Conclusions}
\label{sec:Concls}

We have studied the predictions of the $Z_2$ symmetric 2HD models near the decoupling limit, motivated by no discovery so far of any new scalar. After integrating out the heavy particles, the models can be formulated in terms of  effective Lagrangians that depend on only three parameters: $\tan\beta$, the (approximately) common  mass of the heavy scalars $\tm_2$, and a complex coupling $\tgl_6$ in the scalar potential.  Due to the small number of free parameters, the effective models predict many correlations between observables, which are relevant both for analysing the present experimental constraints and the future discovery prospects.

The effective 2HD models are interesting theoretical laboratories for investigating the interplay of the flavour and collider constraints on the extensions of the SM, along the lines proposed in Refs.~\cite{Alonso-Gonzalez:2021jsa,Alonso-Gonzalez:2021tpo}. The flavour $Z_2$ symmetry eliminates the tree-level contributions to the FCNC and, therefore, new  scalars with a relatively low mass scale may couple to the SM gauge bosons and fermions consistently with the constraints from the very high precision flavour data.  In the effective approach, the predicted contributions to flavour-changing observables depend only on the value of $\tan\beta$ (entering through Yukawa couplings) and on the degenerate heavy scalar masses $\tm_2$. A fit to almost 60 observables gives the exclusions limits on $\tm_2$, depending on type I-IV of the 2HDM and the value of $\tan\beta$: in particular, for the Type IV a lower bound on $\tm_2$ in the range $300-500 \GeV$ is found; a similar limit applies in the Type-II scenario, but only for values of $\tan\beta$ in the range $20-40$, otherwise the bounds are weaker.

With these relatively weak bounds from flavour data, a complementary source of information on the effective Yukawa couplings and on the Lagrangian parameter comes from collider data on the  Higgs production and decays. Strong correlations between different Yukawa couplings are predicted by the models. They have an impact on the results of the fit to the present data and determine the room left for deviations from the SM, important for future discovery prospects. The striking prediction of all four types of 2HD models is that the top quark Yukawa coupling is the least promising to show deviations from the SM value. The correlations among the effective Yukawa couplings are equally important for the bounds on the imaginary parts of the Yukawas inferred from the EDM bounds.

Another prediction following from the fits to flavour and Higgs data is for the expected range of the triple Higgs coupling: the deviations from the SM prediction, in the absence of hierarchies between $\tgl_6$ and $\tgl_{3,4,5}$, are tiny in the Type II and III models for any value of $\tan\beta$ and in the Type IV for large $\tan\beta$; in the other cases the  deviations may be sizeable and therefore testable in the future colliders.

\section*{Acknowledgements}

The authors thanks Francisco Arco García and José Miguel No for useful discussions. A.d.G. and L.M. thanks the Institute of Theoretical Physics and the Faculty of Physics of the University of Warsaw for hospitality during the development of this project. S.P. thanks the Institute of Theoretical Physics of the Universidad Aut\'onoma de Madrid for hospitality during the development of this project.
A.d.G. and L.M. acknowledge partial financial support by the Spanish Research Agency (Agencia Estatal de Investigaci\'on) through the grant IFT Centro de Excelencia Severo Ochoa No CEX2020-001007-S and by the grant PID2019-108892RB-I00 funded by MCIN/AEI/ 10.13039/501100011033, by the European Union's Horizon 2020 research and innovation programme under the Marie Sk\l odowska-Curie grant agreements No 860881-HIDDeN and 101086085-ASYMMETRY. The work of A.d.G. is supported by the European Union's Horizon 2020 Marie Sk\l odowska-Curie grant agreement No 860881-HIDDeN.
The research of S.P. has received partial financial support by the  National Science Centre, Poland, grant DEC-2018/31/B/ST2/02283.

\appendix
\boldmath
\section{From the $Z_2$-original basis to the Higgs basis}
\label{App.A}
\unboldmath
In this appendix, we define the scalar sector of the Lagrangian and its connection to the Higgs basis used in Sec.~\ref{sec:Lag} following the notation of Ref.~\cite{Egana-Ugrinovic:2015vgy}.

The scalar sector is composed of two identical complex scalar fields $\Phi_a $, with $a = 1,2$, which are charged under $SU_L(2) \times U_Y(1)$ and with hypercharge, $Y_{\Phi_a}=1/2$.
The Lagrangian reads
\be
\sL= (D_\mu\Phi_a)^\dagger (D^\mu\Phi_a) +V(\Phi_1,\Phi_2)\, ,
\ee
where the most general renormalizable potential for the two doublets reads
\ba\label{VPhi12}
V(\Phi_1,\Phi_2) =& - m_1^2 (\Phi^\dagger_1\Phi_1) - m_2^2 (\Phi^\dagger_2\Phi_2) - \left[ m_{12}^2 (\Phi^\dagger_1\Phi_2) + \text{h.c.} \right] \\
 &+ \gl_1 (\Phi^\dagger_1\Phi_1)^2 + \gl_2 (\Phi^\dagger_2\Phi_2)^2 + \gl_3 (\Phi^\dagger_1\Phi_1)(\Phi^\dagger_2\Phi_2) + \gl_4 (\Phi^\dagger_1\Phi_2)(\Phi^\dagger_2\Phi_1) \nonumber\\
 &+ \left[ \gl_5 (\Phi^\dagger_1\Phi_2)^2 + \gl_6 (\Phi^\dagger_1\Phi_1)(\Phi^\dagger_1\Phi_2)  + \gl_7 (\Phi^\dagger_2\Phi_2)(\Phi^\dagger_1\Phi_2) + \text{h.c.} \right]\,.
\ea
The parameters $m_1^2, m_2^2, \gl_1, \gl_2, \gl_3$ and $\gl_4$ are real while $m_{12}^2, \gl_5, \gl_6$ and $\gl_7$ are in general complex.

\noindent
In the $\Phi$-basis, both Higgs doublets get a vev. Without loss of generality, one can write
\be
\left<\Phi_1^\dagger\Phi_1\right>=\dfrac{v_1^2}{2}\,,\qquad \left<\Phi_2^\dagger \Phi_2\right>=\dfrac{v_2^2}{2}\,,\qquad \xi\equiv \arg\left<\Phi_1^\dagger\Phi_2\right>
\ee
where, $v_1,v_2,\xi$ are real, $v^2 = v_1^2 + v_2^2 = (246 \, \text{GeV})^2$ and
\be\label{tanbeta}
\tan \beta \equiv \frac{v_2}{v_1} \, .
\ee
 Under the $\mathbb Z_2$ symmetry the doublets transform differently as 
\be
\Phi_1 \to + \Phi_1 \qquad \text{and} \qquad \Phi_2 \to - \Phi_2
\ee
which in turn enforces the condition 
\be
\gl_6 = \gl_7 = 0
\ee
We allow the discrete symmetry to be softly broken in the potential by $m^2_{12}$.

The Higgs basis is defined by the condition that only one doublet admits a vev, i.e.
\be
\left<H_1^\dagger H_1\right>=\dfrac{v^2}{2}\,,\qquad \left<H_2^\dagger H_2\right>=0\,.
\ee
It can be reached by a rotation of the doublets
\be
\label{Hbasis}
\begin{pmatrix}
           e^{-i\xi/2}H_1 \\
           H_2 \\
         \end{pmatrix}
         =
 \begin{pmatrix}
            c_\beta & s_\beta e^{-i\xi} \\ 
          -s_\beta e^{i\xi} & c_\beta \\
         \end{pmatrix}
         \begin{pmatrix}
           \Phi_1\\
           \Phi_2 \\
         \end{pmatrix}
\ee
As the $\mathbb{Z}_2$-symmetry is not defined in this basis, $\gl_6$- and $\gl_7$-like terms are again developed. The potential reads
\begin{align}
    V(H_1,H_2)=\,& \widetilde{m}_1^2 H_1^\dagger H_1 +\widetilde{m}_2^2 H_2^\dagger H_2 -\rlb \widetilde{m}_{12}^2H_1^\dagger H_2 +\text{h.c.}\rrb+\dfrac{1}{2}\widetilde{\gl}_1 \rlb H_1^\dagger H_1\rrb^2+\nn\\
    &+\dfrac{1}{2}\widetilde{\gl}_2 \rlb H_2^\dagger H_2\rrb^2 + \widetilde{\gl}_3 \rlb H_1^\dagger H_1\rrb \rlb H_2^\dagger H_2\rrb + \widetilde{\gl}_4 \rlb H_1^\dagger H_2\rrb \rlb H_2^\dagger H_1\rrb+ \\
    &+\slb \dfrac{1}{2} \widetilde{\gl}_5 \rlb H_1^\dagger H_2\rrb^2  + \widetilde{\gl}_6 \rlb H_1^\dagger H_1\rrb \rlb H_1^\dagger H_2\rrb + \widetilde{\gl}_7 \rlb H_2^\dagger H_2\rrb \rlb H_1^\dagger H_2\rrb +\text{h.c.} \srb\,.
\end{align}
The tilded parameters are defined in terms of the original ones 
via
\ba
\label{wtm-m}
\widetilde m_1^2 &= m_1^2 c_\beta^2 + m_2^2 s_\beta^2 + s_{2\beta}  \Re(m_{12}^2 e^{i\xi})\,,\\
\widetilde m_2^2 &= m_1^2 s_\beta^2 + m_2^2 c_\beta^2 -  s_{2\beta} \Re(m_{12}^2 e^{i\xi})\,,\\
\widetilde m_{12}^2 e^{i\xi/2} &= (m_2^2 - m_1^2)c_\beta s_\beta + \Re(m_{12}^2 e^{i\xi})c_{2\beta} + i \Im(m_{12}^2 e^{i\xi})\,,
\ea
and 
\ba
\label{wtl-l}
\widetilde \gl_1 &=  \gl_1 c_\beta^4 + \gl_2 s_\beta^4 + \gl_{345}c^2_\beta s^2_\beta \,,\\
\widetilde \gl_2 &=  \gl_1 s_\beta^4 + \gl_2 c_\beta^4 + \gl_{345}c^2_\beta s^2_\beta \,,\\
\widetilde \gl_3 &=  \gl_3 + ( \gl_1 + \gl_2 - \gl_{345}) \frac{s^2_{2\beta} }{4} \,,\\
\widetilde \gl_4 &=  \gl_4 + ( \gl_1 + \gl_2 - \gl_{345}) \frac{s^2_{2\beta} }{4} \,,\\
\widetilde \gl_5e^{i\xi/2} &=   \dfrac{s^2_{2\beta} }{4}( \gl_1 + \gl_2 - \gl_{345}) +\Re(\gl_5e^{2i\xi})+i c_{2\beta}\Im(\gl_5e^{2i\xi})  \,,\\
\widetilde \gl_6e^{i\xi/2} &=  -\frac{s_{2\beta} }{2}( \gl_1 c_\beta^2 - \gl_2 s_\beta^2 - \gl_{345} c_{2\beta} - i \Im(\gl_5e^{2i\xi}))   \,,\\
\widetilde \gl_7e^{i\xi/2} &= -\frac{s_{2\beta} }{2} ( \gl_1 s_\beta^2 - \gl_2 c_\beta^2 + \gl_{345} c_{2\beta} + i \Im(\gl_5e^{2i\xi})) 
\ea
where $\gl_{345} = \gl_3 + \gl_4 + \Re(\gl_5e^{2i\xi})$.

\footnotesize

\bibliographystyle{BiblioStyle}
\bibliography{Bibliography}

\end{document}